\documentclass{article}

\usepackage{graphicx,subfigure,psfrag,dsfont,amssymb,amsthm,amsmath,a4wide}

\psfrag{beta}{$\beta$}
\psfrag{eta}{$\eta$}
\psfrag{sigma}{$\sigma$}
\psfrag{sigma_min}{$\sigma_{\rm min}$}
\psfrag{sigma_max}{$\sigma_{\rm max}$}
\psfrag{beta_star}{$\beta^*$}
\psfrag{eta-}{$\eta-1$}
\psfrag{eta+}{$\eta+1$}
\psfrag{em}{$m$}

\newtheorem{theorem}{Theorem}

\newtheorem{proposition}{Proposition}
\newtheorem{corollary}{Corollary}

\newcommand{\nr}{n}
\DeclareMathOperator*{\argmax}{argmax}

\graphicspath{{Images/}}

\begin{document}

\title{Optimal Tradeoff between Exposed and Hidden Nodes in Large Wireless Networks}
\author{P.M. van de Ven$^{1,2}$ \and A.J.E.M. Janssen$^{2,3}$ \and J.S.H. van Leeuwaarden$^{1,2}$}

\footnotetext[1]{Eindhoven University of Technology, Department of Mathematics and Computer Science, P.O. Box 513, 5600 MB Eindhoven, The Netherlands}
\footnotetext[2]{Eurandom, P.O. Box 513, 5600 MB Eindhoven, The Netherlands}
\footnotetext[3]{Eindhoven University of Technology, Department of Electrical Engineering, P.O. Box 513, 5600 MB Eindhoven, The Netherlands}

\maketitle

\begin{abstract}
Wireless networks equipped with the CSMA protocol are subject to collisions due to interference. For a given interference range we investigate the tradeoff between collisions (hidden nodes) and unused capacity (exposed nodes). We show that the sensing range that maximizes throughput critically depends on the activation rate of nodes. For infinite line networks, we prove the existence of a threshold: When the activation rate is below this threshold the optimal sensing range is small (to maximize spatial reuse). When the activation rate is above the threshold the optimal sensing range is just large enough to preclude all collisions. Simulations suggest that this threshold policy extends to more complex linear and non-linear topologies.
\end{abstract}


\section{Introduction}

Carrier sense multiple-access (CSMA) type protocols form a popular class of medium access protocols for wireless networks. The first CSMA protocol was introduced by Kleinrock and Tobagi~\cite{KlTo75} in 1975, and has seen many incarnations since, including the widely used 802.11 standard. In this paper we provide an asymptotic analysis of large wireless networks operating under CSMA, in the presence of collisions.

CSMA is a randomized protocol that allows nodes to access the medium in a distributed manner. The absence of a centralized scheduler creates more flexibility and allows for the deployment of larger networks. An early example of such a randomized procedure is the ALOHA protocol \cite{Abramson70}, which forces nodes to wait for some random backoff period before starting a transmission, in order to reduce the likelihood of nearby nodes transmitting simultaneously. The latter event would cause the signals to interfere with each other, and may result in a collision that renders the transmissions useless. CSMA improves upon ALOHA by letting nodes {\it sense} their surroundings to detect the presence of other transmitting nodes. If a node detects at least one active (i.e.~transmitting) node within its sensing range, its backoff timer is frozen, deferring the countdown until the channel is sensed clear. Using this mechanism, collisions can be further reduced.

A key performance measure in wireless networks is throughput, which we define as the average number of successful transmissions per unit of time. We investigate the relation between the sensing range and the throughput. The effect of the sensing range can be understood as follows. A small sensing range allows for more simultaneous transmissions, but is less effective in reducing collisions. On the other hand, a large sensing range admits fewer transmissions, but also mitigates interference.
The main contribution of this paper is the examination of this tradeoff in relation to its effect on the throughput.

The network is characterized by the sensing range and the interference range. A node can only initiate a new transmission when all nodes within its sensing range are inactive. This transmission is successful when all nodes within the interference range of the destination node are inactive, and fails otherwise. The network performance suffers from two complementary issues: hidden nodes and exposed nodes (see \cite{ToKl75}). Hidden nodes are nodes located outside the sensing range of the transmitter and are therefore not detected by the carrier-sensing mechanism. Hidden nodes cause collisions as they are within the receiver's interference range. Exposed nodes are nodes located outside the receiver's interference range but inside the sender's sensing range. So despite being harmless to the transmission, exposed nodes are nevertheless blocked. As the sensing range grows, the number of hidden nodes decreases, and the number of exposed nodes increases.

In recent years the carrier-sensing tradeoff between hidden and exposed nodes has received much attention \cite{LiHo07,MaViRoZh09,YaVa05,ZhFa06}. Most of these analytic studies make the assumption that the activity of nodes and their backoff processes are independent, which greatly simplifies the analysis. The interaction between nodes, however, should be taken into account, as it is typical for the distributed control and has a large impact on the performance of the network. We do take into account this interaction, by keeping track of the activity of nodes over time. The classical model for such interaction in wireless networks is developed in Boorstyn and Kershenbaum~\cite{BoKe80}. This model has been used in recent years to study throughput-optimality~\cite{RaShSh09} and fairness~\cite{DeBoVeHi08,DuDoTh07,WaKa05,VeLeDeJa09} in a setting without collisions. The stability region for large wireless networks with collisions was investigated in \cite{BoMcPr08}.

In the spirit of \cite{BoKe80}, we model the network as a continuous-time Markov process with interaction between the nodes, so that nodes within a certain distance of an active node are silenced, just as in CSMA. Such interaction is referred to in statistical physics as {\it hard-core} interaction. This paper is part of a larger program to study wireless networks via hard-core models from statistical physics. Typical for such models is the existence of a Gibbs measure that describes the stationary distribution. This Gibbs measure is normalized by the partition function, which involves a computationally cumbersome summation over all possible configurations. A substantial ingredient of this paper is to characterize and approximate the partition function. We shall consider the network, and thus the partition function, in the asymptotic regime where the number of nodes in the network tends to infinity. For such infinite line networks we are able to obtain structural results on the joint effect of hidden nodes and exposed nodes.We determine analytically the throughput-optimal sensing range that achieves the best tradeoff between reducing hidden nodes and preventing exposed nodes.


The remainder of this paper is structured as follows. In Section~\ref{sec:model_auxiliary} we introduce the model, and derive some auxiliary results. Section~\ref{sec:main_results}  discusses the main results on the carrier-sensing tradeoff. In Section~\ref{sec:detailed_study} we perform a detailed study of the partition function. In Section~\ref{sec:discussion} we validate the analytical results for the line network by simulation, and we investigate networks with more general topologies. In Section~\ref{sec:proofs} we present the proofs of those results that are not already proved in earlier sections.

\section{Model description}\label{sec:model_auxiliary}

We consider a linear array of $2\nr+1$ nodes, and we denote the set of all nodes by $\mathcal{N} = \{-\nr,\dots,\nr\}$. Whenever a node activates, it transmits a single packet to a neighboring node. With probability $\psi$, the packet is intended for its right neighbor, and with probability $1 - \psi$ for its left neighbor. To accommodate this, we introduce (pure destination) nodes~$\nr+1$ and ~$-(\nr+1)$, which receive packets, but do not transmit packets themselves. As will be shown in Proposition~\ref{pro:throughput}, the throughput is insensitive to the parameter $\psi$. We assume that all nodes are saturated, meaning that they have an infinite supply of packets available.

After each transmission nodes enter a backoff period, meaning that they will remain inactive for some time. The length of the backoff period is assumed to be exponentially distributed with mean~$1/\sigma$. We assume all nodes to have the same sensing range~$\beta$, so that node~$v$ is prohibited from transmitting whenever at least one node $w$ for which $|v-w| \leq \beta$ is active (i.e.~transmitting), in which case we say that node $v$ is {\it blocked} by node $w$. So when a node finishes its backoff period and it finds at least one node within distance $\beta$ active, it enters a new backoff period. When a node finds all nodes within distance~$\beta$ inactive upon finishing backoff, it starts a transmission.
Transmissions last for an exponentially distributed duration with unit mean. Under these assumptions, the $(2 \nr + 1)$-dimensional process that describes the activity of nodes is a continuous-time Markov process. Each state of the Markov process is described by
\begin{equation}
\omega = (\omega_{-\nr},\dots,\omega_{\nr}) \in \{0,1\}^{2n+1},
\end{equation}
where $\omega_v = 1$ when node~$v$ is active, and $\omega_v = 0$ otherwise. Let $\Omega \subseteq \{0,1\}^{2n+1}$ be the set of all {\em feasible} states.
Here we call $\omega$ feasible if no two $1$'s in $\omega$ are $\beta$ positions or less apart, i.e.,~$\omega_v\omega_w=0$ if  $1 \le |v-w|\leq \beta$. Let $e_v$ denote the vector with all zeros, except for a 1 at position $v$.
The Markov process that describes the activity of nodes is then fully specified by the state space $\Omega$ and the transition rates
\begin{equation}\label{eqn:trans_rates}
r(\omega,\omega')=\left\{
                    \begin{array}{ll}
                      \sigma & \hbox{if $\omega'=\omega + e_v$,} \\
                      1 & \hbox{if $\omega'=\omega - e_v$,} \\
                      0 & \hbox{otherwise}.
                    \end{array}
                  \right.
\end{equation}
It is well known that this is a reversible Markov process (see \cite{BoKe80,PiYe86}) with limiting distribution
\begin{equation}\label{eqn:lim_dist}
\pi(\omega)=\left\{
         \begin{array}{ll}
           Z_{2n+1}^{-1}\prod_{v=-n}^{n}\sigma^{\omega_v} & \hbox{if $\omega$ is feasible,} \\
           0 & \hbox{otherwise,}
         \end{array}
       \right.
\end{equation}
with $Z_{2n+1}$ the partition function or normalization constant of the probability distribution $\pi$. The partition function can be defined recursively as (see \cite{BoKe80,PiYe86})
\begin{equation}\label{partrec}
Z_i =
\left\{
  \begin{array}{ll}
    1 + i \sigma &                    \hbox{$i = 0,1,\dots,\beta+1$}, \\
    Z_{i-1} + \sigma Z_{i-\beta-1}  & \hbox{$i \ge \beta+2$}.
  \end{array}
\right.
\end{equation}
The sequence $(Z_i)_{i=0}^\infty$ is well studied. In fact, for a network with $i$ nodes, $Z_i$ represents the partition function, defined as the summation of probability over all possible states. Straightforward calculations show that the
the generating function $G_Z(x)$ of $Z_i$ can be written as (see e.g.~Pinksy and Yemini~\cite{PiYe86})
\begin{equation}\label{eqn:generating_function}
G_Z(x) = \sum_{i = 0}^\infty Z_i x^i  = \frac{x-1 + \sigma x^{\beta+1} - \sigma x}{(x-1) (1 - x - \sigma x^{\beta+1})}.
\end{equation}
Let $\lambda_0,\ldots,\lambda_\beta$ denote the $\beta+1$ distinct (see Proposition~\ref{pro:distinct}) roots of
\begin{equation}\label{eqn:equation_lambda}
\lambda^{\beta+1} - \lambda^\beta - \sigma = 0.
\end{equation}
We denote by $\lambda_0$ the unique positive real root for which $\lambda_0 > |\lambda_j|,\ j \neq 0$ (see~\cite{PiYe86}). Applying partial fraction expansion to \eqref{eqn:generating_function} yields the following result (proved in Section~\ref{sec:proofs}):
\begin{proposition}\label{pro:fraction_expansion}
The partition function $Z_i$ is given by
\begin{equation}\label{dfg}
Z_i = \sum_{j = 0}^\beta c_j \lambda_j^i \quad , i =0,1,\dots,
\end{equation}
where $\lambda_j$ are the roots of \eqref{eqn:equation_lambda}, and
\begin{equation}\label{eqn:c_j}
c_j = \frac{\lambda_j^{\beta + 1}}{(\beta+1)\lambda_j - \beta}.
\end{equation}
\end{proposition}

\begin{figure}[h]
 \begin{center}
 \subfigure[\small{Node~3 is a hidden node, and may interfere with the transmission between nodes~0 and~1.}]{\label{fig:hidden} \includegraphics[height = 0.15\textheight]{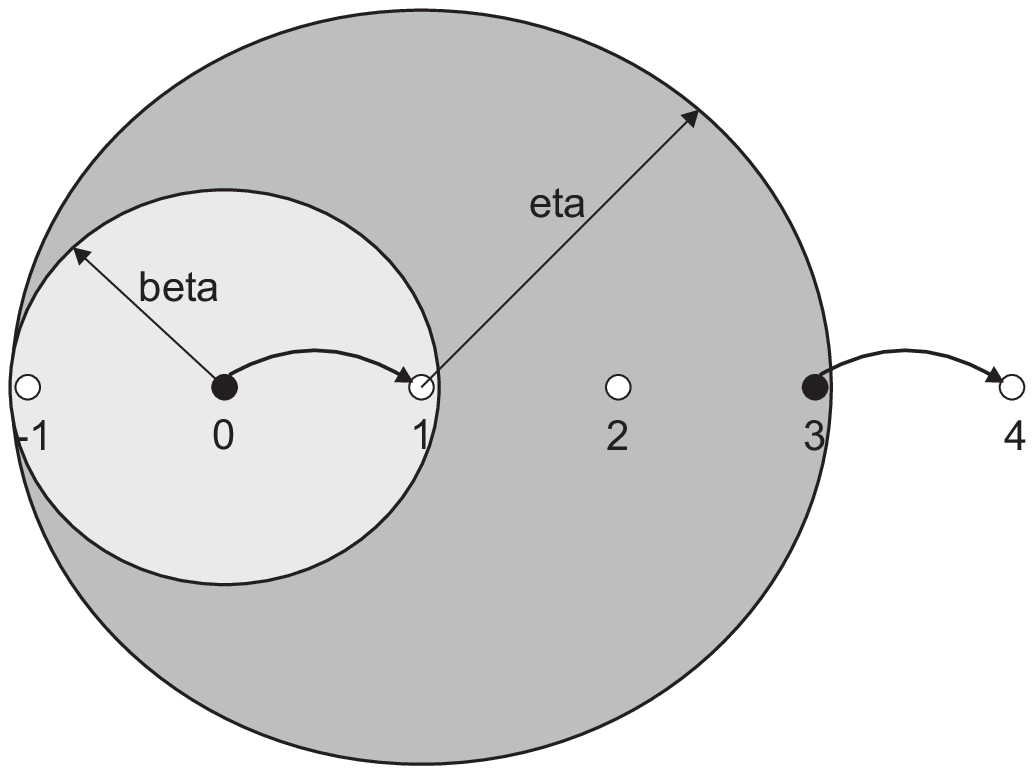}}\hspace{0.5cm}
 \subfigure[\small{Node~0 is an exposed node, unnecessarily silenced by the transmission between nodes~2 and~3.}]{\label{fig:exposed} \includegraphics[height = 0.15\textheight]{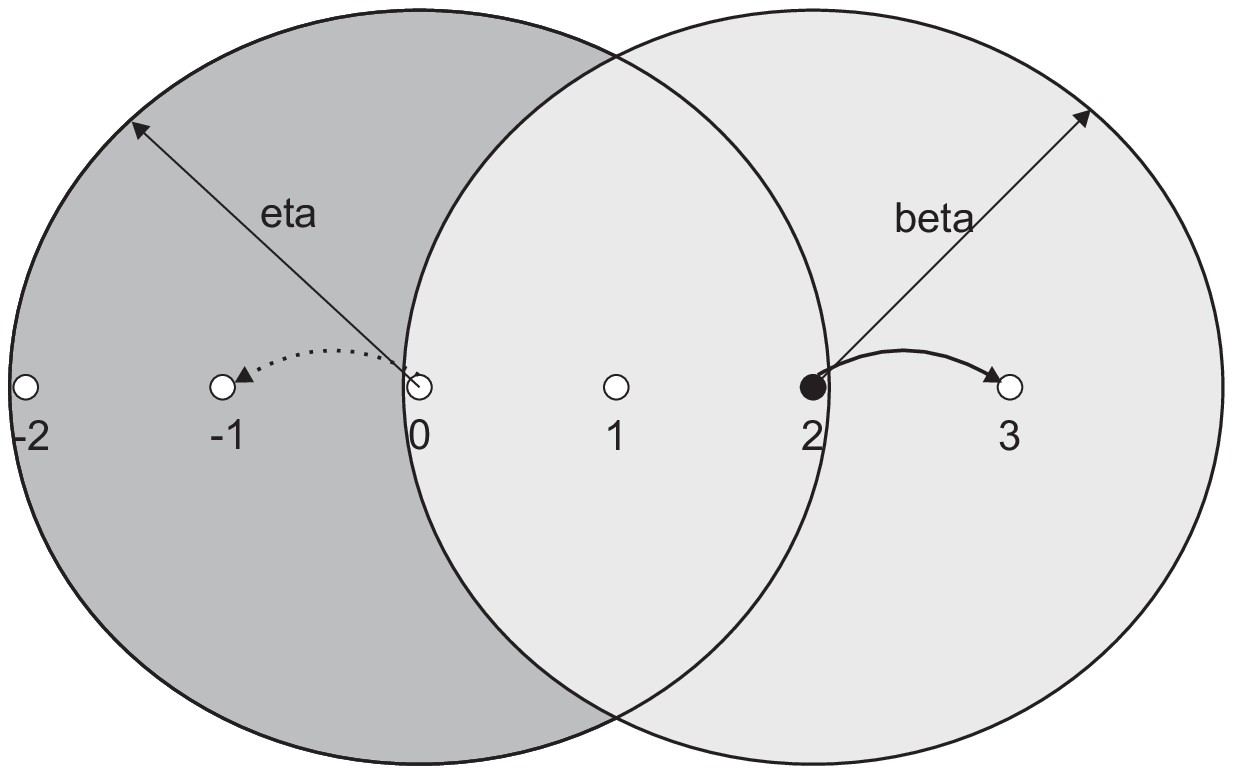}}
 \end{center}
 \caption{Examples of hidden and exposed nodes.}
 \label{fig:hidden_exposed}
\end{figure}

To model interference, we introduce an interference range~$\eta$.
A transmission succeeds if and only if at the start of this transmission no nodes within distance $\eta$ of the receiving node are already active. This type of interference is referred to in the literature as the {\it perfect capture} collision model \cite{BoKe80}.
Note that neither~\eqref{eqn:trans_rates} nor~\eqref{eqn:lim_dist} depends on $\eta$, as collisions have no impact on the dynamics of the system. Using the sensing range $\beta$ and interference range $\eta$ we can define formally hidden nodes and exposed nodes. Consider a transmission from node~$v$ to node~$w$. Hidden nodes are then defined as nodes that are outside the sensing range of~$v$, but within the interference range of~$w$. Such nodes are not blocked by the activity of node~$v$, but their proximity to node~$w$ makes the hidden nodes harmful to the transmission from~$v$ to~$w$. Conversely, exposed nodes are those nodes that are within the sensing range of $v$, but outside the interference range of $w$. Such nodes are blocked by an ongoing transmission from $v$ to $w$, despite the fact that they will not cause this transmission to fail. Denote by $\mathcal{H}_r$ ($\mathcal{H}_l$) the set of hidden nodes of transmissions from node~0 to node~1 (node~-1): all nodes  outside the sensing range of 0, but within the interference range of the receiving node~1 (node~-1). By $\mathcal{E}_r$ ($\mathcal{E}_l$) we denote the set of nodes to which this transmission is exposed, so all nodes  within the sensing range of~0, but outside the interference range of the receiving node. For completeness we let $\mathcal{B}_r$ ($\mathcal{B}_l$) denote the set of all remaining nodes that block transmissions from node~$0$ to node~$1$ (node~-1). This yields:
\begin{align*}
\mathcal{H}_r &= \big\{\, v \in \mathcal{N}\ \big|\ |v| \ge \beta + 1, \ |v-1| \le \eta \,\big\}, \quad \mathcal{H}_l = \big\{\, v \in \mathcal{N}\ \big|\ |v| \ge \beta + 1, \ |v + 1| \le \eta \,\big\},\\
\mathcal{E}_r &= \big\{\, v \in \mathcal{N}\ \big|\ |v| \le \beta, \ |v - 1| \ge \eta + 1 \,\big\}, \quad \mathcal{E}_l = \big\{\, v \in \mathcal{N}\ \big|\ |v| \le \beta, \ |v + 1| \ge \eta + 1 \,\big\},\\
\mathcal{B}_r &= \big\{\, v \in \mathcal{N}\ \big|\ |v| \le \beta, \ |v - 1| \le \eta \,\big\}, \quad \mathcal{B}_l = \big\{\, v \in \mathcal{N}\ \big|\ |v| \le \beta, \ |v + 1| \le \eta \,\big\}.
\end{align*}
So $\mathcal{E}_r \cup \mathcal{B}_r = \mathcal{E}_l \cup \mathcal{B}_l = \big\{v \in \mathcal{N}\ \big|\ |v| \le \beta \big\}$. An example is given in Figure~\ref{fig:hidden}. Node 3 is a hidden node, as it interferes with the transmission from node $0$ to node $1$ ($\eta=2$) despite the carrier-sensing mechanism ($\beta=1$). In Figure~\ref{fig:exposed} node 0 is an exposed node to the transmission from node~2 to node~3 because it would not interfere ($\eta=2$) with this transmission but is nevertheless silenced by the activity of node~2 ($\beta=2$).

We focus on node~0 (the node in the middle of the network) and in particular its throughput $\theta_n(\beta,\eta,\sigma)$ defined as the average number of successful transmissions per unit of time.
\begin{proposition}\label{pro:throughput}
The throughput of node~0 is given by
\begin{align}\label{eqn:throughput}
\theta_n(\beta,\eta,\sigma) = \sigma \frac{Z_{\nr-\max\{\beta,\eta-1\}} Z_{\nr-\max\{\beta,\eta+1\}}}{Z_{2\nr+1}}.
\end{align}
\end{proposition}

\begin{proof}
Denote by $\theta_r$ $(\theta_l)$ the rate of successful transmission of node~0 to node~1 (node~-1), so $\theta_n(\beta,\eta,\sigma) = \theta_r + \theta_l$. The activation attempts to node 1 (node -1) occur according to a Poisson process with rate $\sigma \psi$ (rate $\sigma (1 - \psi)$). We first consider activation attempts to node 1. Whether an activation attempt is successful depends on the state of the system when this attempt occurs. Define
\begin{align*}\label{eqn:def_hidden_exposed}
A_1 &= \big\{\, \omega \in \Omega\ \big|\ \exists{ v\in \mathcal{B}_r \cup \mathcal{E}_r}: \omega_v = 1\,\big\},\\
A_2 &= \big\{\, \omega \in \Omega\ \big|\ \forall{ v\in \mathcal{B}_r \cup \mathcal{E}_r}: \omega_v = 0,\ \exists{v \in \mathcal{H}_r}:\omega_v = 1\,\big\},\\
A_3 &= \big\{\, \omega \in \Omega\ \big|\ \forall{ v\in \mathcal{B}_r \cup \mathcal{E}_r \cup \mathcal{H}_r}: \omega_v = 0\,\big\}.
\end{align*}
When the system is in state $\omega \in A_1$, the attempt is blocked and node~0 remains in its current state. When the system is in a state $\omega \in A_2$, node~0 is not blocked so it activates. However, at least one hidden node is active so the transmission fails and does not contribute to the throughput. When the system is in state $\omega \in A_3$, the perfect capture assumption guarantees a successful transmission. It follows from the PASTA property (cf.~\cite{Asmussen03}) that the probability of an arbitrary activation attempt resulting in a successful transmission is equal to the limiting probability of the system being in a state $\omega \in A_3$. So the rate of successful transmissions initialized (and thus the throughput) is given by
\begin{equation}\label{eqn:proof_throughput_1}
\theta_r = \sigma \psi \sum_{\omega \in A_3} \pi(\omega).
\end{equation}
From the definitions of $\mathcal{B}_r$, $\mathcal{E}_r$ and $\mathcal{H}_r$ we see that
\begin{equation}
A_3= \big\{\, \omega \in \Omega\ \big|\ \forall{ v\in (D_1 \cup D_2)^c}: \omega_v = 0\,\big\}, \label{eqn:proof_throughput_1b}
\end{equation}
where
\begin{equation}
D_1 = \{-\nr,\dots,-\max\{\beta,\eta-1\}-1\}, \quad D_2 = \{\max\{\beta,\eta+1\}+1,\dots,\nr\}.
\end{equation}

Let $Z_D$ denote the partition function for a subset of nodes $D \subseteq \mathcal{N}$ defined as \newline $Z_D = \sum_{\omega \in \Omega,\ \forall v\in D^c: \omega_v = 0} \prod_{v = -n}^{n} \sigma^{\omega_v}$. Then
\begin{equation}\label{eqn:proof_throughput_3}
\theta_r = \sigma \psi \frac{Z_{D_1 \cup D_2}}{Z_\mathcal{N}}.
\end{equation}
The model on the line has the property that by conditioning on the activity of one of the nodes, its state space can be decomposed, leading to two smaller instances of the same model on the line. In particular, we know that $Z_{D_1 \cup D_2} = Z_{D_1} Z_{D_2}$ (see \cite[Equation (15)]{BoKe80}), so that
\begin{equation}
\theta_r = \sigma \psi \frac{Z_{D_1} Z_{D_2}}{Z_{\mathcal{N}}} = \sigma \psi \frac{Z_{\nr-\max\{\beta,\eta-1\}}  Z_{\nr-\max\{\beta,\eta+1\}}}{Z_{2\nr+1}},
\end{equation}
where $Z_i$ denotes the partition function of a network with $i$ consecutive nodes on a line.
Similarly,
\begin{equation}\label{eqn:proof_throughput_5}
\theta_l = \sigma (1- \psi) \frac{Z_{\nr-\max\{\beta,\eta-1\}}  Z_{\nr-\max\{\beta,\eta+1\}}}{Z_{2\nr+1}}.
\end{equation}
and~\eqref{eqn:throughput} follows by adding $\theta_r$ and $\theta_l$.
\end{proof}

\section{Main results}\label{sec:main_results}

Our principal aim is to choose the sensing range $\beta$ so that the throughput $\theta_n(\beta,\eta,\sigma)$ is maximized for a given $\eta$ and $\sigma$. Define
\begin{equation}
\beta^*_n = \argmax_\beta \theta_n(\beta,\eta,\sigma).
\end{equation}
Determining $\beta^*_n$ corresponds to quantifying and optimizing the tradeoff between preventing collisions through interference (preventing hidden nodes by setting $\beta$ large) and allowing harmless transmissions (preventing exposed nodes by setting $\beta$ small). We want to obtain structural insights in how to choose $\beta^*_n$, and for this purpose the expressions for $Z_i$ in \eqref{dfg} and $\theta_n(\beta,\eta,\sigma)$ in \eqref{eqn:throughput} are too cumbersome. Therefore, we investigate the throughput in the regime where the network becomes large ($n \rightarrow \infty$), so that \eqref{eqn:throughput} simplifies considerably, allowing for more explicit analysis.
The analytic results that we obtain for the infinite network provide remarkably sharp approximations for the finite network; see Section \ref{sec:finite}. All proofs that are not given in this section are provided in Section~\ref{sec:proofs}.

We start by presenting the limiting expression for $\theta_n(\beta,\eta,\sigma)$ as the size of the network becomes infinite:
\begin{proposition}
Let $\lambda_0$ denote the unique positive real root of \eqref{eqn:equation_lambda}. Then
\begin{equation}\label{eqn:asymptotic_throughput}
\theta(\beta,\eta,\sigma) = \lim_{n \rightarrow \infty} \theta_n(\beta,\eta,\sigma) = \sigma \frac{\lambda_0^{\beta-f(\beta)}}{(\beta+1) \lambda_0 - \beta},
\end{equation}
where
\begin{equation}\label{eqn:f(beta)}
f(\beta) = \left\{
\begin{array}{ll}
2 \eta  &   {\rm if~} 0 \le \beta \le \eta-1,\\
\eta + \beta + 1    &   {\rm if~} \eta-1 \le \beta \le \eta + 1,\\
2 \beta &   {\rm if~} \beta \ge \eta+1.
\end{array} \right.
\end{equation}
\end{proposition}

\begin{proof}
From Rouch\'{e}'s theorem (see De Bruijn \cite{DeBruijn81}) it readily follows  that $\lambda_0 > |\lambda_j|$ for $j = 1,\dots,\beta$, and so from \eqref{dfg} we get
\begin{equation}
Z_i = c_0 \lambda_0^i\left(1 + o(1)\right), \quad i \rightarrow \infty.
\end{equation}
Hence
\begin{equation}
 \lim_{n \rightarrow \infty} \theta_n(\beta,\eta,\sigma) = \lim_{n \rightarrow \infty} \sigma \frac{c_0 \lambda_0^{n - \max\{\beta,\eta-1\}} c_0 \lambda_0^{n - \max\{\beta,\eta + 1\}}}{c_0 \lambda_0^{2 n +1}} = \sigma c_0 \lambda_0^{- \max\{\beta,\eta - 1\} - \max\{\beta,\eta+1\} - 1},
\end{equation}
which yields \eqref{eqn:f(beta)}.
\end{proof}

Now that we have the limiting expression for the throughput in \eqref{eqn:asymptotic_throughput} we opt for an asymptotic analysis. That is, instead of searching for $\beta^*_n$, we shall search for its asymptotic counterpart
\begin{equation}
\beta^* = \argmax_\beta \theta(\beta,\eta,\sigma),
\end{equation}
where we henceforth consider $\theta$ as a continuous function of the real variable $\beta \ge 0$.
In Section \ref{sec:finite} we show that the errors $|\theta_n-\theta|$ and $|\beta^*_n-\beta^*|$ become small, already for moderate values of $n$. Because we consider from here onwards an infinite line of nodes, all nodes have the same number of nodes within their sensing range. This removes all boundary effects, and all nodes have the same throughput, which is why just investigating node~0 is sufficient to investigate the entire network.

\begin{proposition}\label{thminterbeta} $\beta^* \in [\eta-1,\eta+1]$.
\end{proposition}

The result of Proposition~\ref{thminterbeta} can be understood as follows. By increasing $\beta$ beyond $\eta+1$, no additional collisions are prevented, but an increasing number of nodes is silenced. On the other hand, the nodes that become unblocked when decreasing $\beta$ below $\eta-1$, cause collisions when they activate. Although this result may seem intuitively clear, to the authors' knowledge such a result has not been proved rigourously (at least not in the present setting). Note that for all values $\beta \in [\eta-1,\eta+1]$, we can rewrite \eqref{eqn:asymptotic_throughput} as
\begin{equation}\label{eqn:throughput_alternative}
\theta(\beta,\eta,\sigma)  = g(\beta) \cdot \frac{(\lambda_0(\beta))^{\beta-\eta-1}}{\beta+1}
\end{equation}
with
\begin{equation}\label{eqn:g}
g(\beta) = \frac{\lambda_0(\beta) - 1}{\lambda_0(\beta) - \frac{\beta}{\beta+1}}\rightarrow 1, \quad \beta\rightarrow\infty.
\end{equation}

We are now in the position to present our main result. While we already know that the optimal sensing range is contained in the interval $[\eta-1,\eta+1]$, the next result is more specific.
\begin{theorem}\label{thm:threshold_interval}
There exists a threshold interval $[\sigma_{\rm min},\sigma_{\rm max}]$ such that
\begin{equation}
\beta^* = \left\{
\begin{array}{ll}
\eta-1  &   \mathrm{if~} \sigma \le \sigma_{\rm min},\\
\eta+1  &   \mathrm{if~} \sigma \ge \sigma_{\rm max},
\end{array}\right.
\end{equation}
and  $\beta^*$ increases from $\eta- 1$ to $\eta + 1$ when $\sigma$ increases from $\sigma_{\rm min}$ to $\sigma_{\rm max}$.
\end{theorem}

The proof of Theorem~\ref{thm:threshold_interval}, see Section~\ref{sec:proofs}, follows from a detailed study of $\theta(\beta,\eta,\sigma)$ which involves implicit differentiation with respect to $\beta$ (since $\lambda_0(\beta)$ is defined implicitly).

\begin{figure}[h]
 \begin{center}
    \includegraphics[width = 0.7\linewidth]{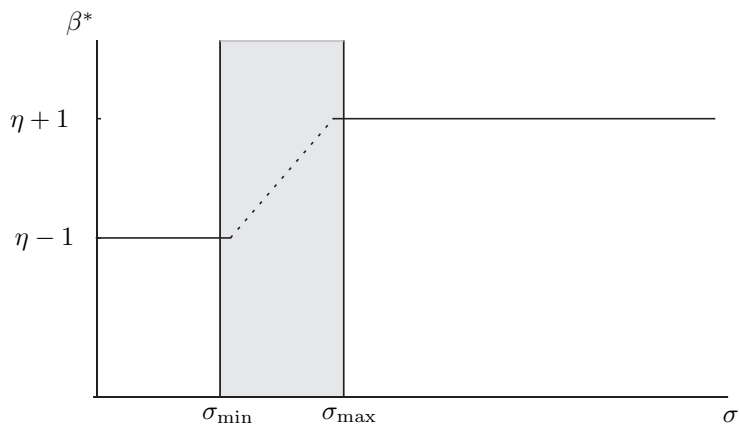}
 \end{center}
 \caption{The optimal sensing range $\beta^*$ as a function of $\sigma$.}
 \label{fig:beta^*}
\end{figure}

Theorem \ref{thm:threshold_interval} can be interpreted as follows (see Figure~\ref{fig:beta^*}).
When $\sigma$ is large, nodes activate very quickly after finishing their previous transmissions. In the language of statistical physics, the system temperature decreases, and the system typically gets stuck in maximal independent sets of active nodes (the configurations with the highest energy level).
When the system is in a maximal independent set, and if collisions are not ruled out, an activating node suffers a collision almost surely. This explains why for $\sigma$ large, the optimal sensing range is $\beta  = \eta +1$, preventing collisions completely. On the other hand, when $\sigma$ is small, collisions become rare, as few nodes are active simultaneously. In this case, the throughput is best served by increasing the spatial reuse, that is, decreasing the sensing range (up to $\eta - 1$). This explains the result of Theorem~\ref{thm:threshold_interval} for $\sigma$ small.

Note that Theorem~\ref{thm:threshold_interval} does not give the exact values of $\sigma_{\rm min}$ and $\sigma_{\rm max}$. Instead, we give below an estimate of the location and width of the threshold interval.
\begin{theorem}\label{thm:length_threshold_interval}
Let $\kappa = \frac{\tau}{\eta+1}$ with $\tau = (\sqrt{5} - 1)/2$. \\
(i) The threshold interval is bounded as
\begin{equation}\label{eqn:threshold}
[\sigma_{\rm min},\sigma_{\rm max}] \subseteq [\kappa(1 + \kappa)^{\eta-1},\kappa(1 + \kappa)^{\eta+1}].
\end{equation}
(ii) The width of the threshold interval is asymptotically given as
\begin{equation}
\sigma_{\rm max} - \sigma_{\rm min} \sim \frac{2 {\rm e}^{\tau}}{7 + 4 \tau} \left( \frac{1}{\eta+1}\right)^2 \quad {\rm as~} \eta \rightarrow \infty.
\end{equation}
\end{theorem}

Here we say that $f(\eta) \sim g(\eta)$ if $f(\eta)/g(\eta) \rightarrow 1$ as $\eta \rightarrow \infty$. From Theorem~\ref{thm:length_threshold_interval}(ii) we see that the width of the threshold interval is $\mathcal{O}(\eta^{-2})$. Therefore, the interval width decreases rapidly as a function of $\eta$, and we can speak of an almost immediate transition from one regime ($\beta^*=\eta-1$) to the other ($\beta^*=\eta+1$). As a by-product of the proof of Theorem~\ref{thm:length_threshold_interval}(ii) we obtain sharp approximations for $\sigma_{\rm min}$ and $\sigma_{\rm max}$, see \eqref{eqn:approximation_sigmas}-\eqref{eqn:equation_p11}:
\begin{equation}\label{eqn:approximations_sigma_min_max}
\hat{\sigma}_{\rm min} = \hat{\mu}_- (1 + \hat{\mu}_-)^{\eta - 1}, \quad \hat{\sigma}_{\rm max} = \hat{\mu}_+ (1 + \hat{\mu}_+)^{\eta + 1},
\end{equation}
with $\hat{\mu}_{\pm} = \tau/(\eta + \alpha_{\pm})$ and $\alpha_{\pm}$ given as $\alpha$ in~\eqref{eqn:equation_p11} with $\gamma = \pm 1$.

\subsection{Throughput limiting behavior}

We now consider some limiting regimes for which we can make more explicit statements about the throughput. From Theorem~\ref{thm:length_threshold_interval} we can already see that the threshold interval moves in the direction of zero as $\eta$ becomes large which implies that $\beta^*=\eta+1$ for small values of $\sigma$. The next result shows that in the regime where $\eta$ becomes large, the maximum throughput tends to zero.
\begin{proposition}\label{prop:lim:throughput}
Let $\sigma > 0$ be fixed. As $\eta \rightarrow \infty$,
\begin{equation}
\max_{\beta} \theta(\beta,\eta,\sigma) = \frac{1}{\eta+2}\left(1 + \mathcal{O}\left(\frac{1}{\ln(\eta+1)}\right)\right).
\end{equation}
\end{proposition}

For $\beta \ge \eta+1$ our model reduces to a model without collisions that was studied extensively in \cite{BoKe80,PiYe86,Ba2004,ZaMo06,DuDoTh09,VeLeDeJa09}.
In particular, one immediately obtains from \eqref{eqn:asymptotic_throughput} the following result:
\begin{corollary}\label{col:limiting_throughput_beta_large}
Let $\beta \ge \eta+1$. Then
\begin{equation}\label{sgsd}
\theta(\beta,\eta,\sigma) = \frac{\lambda_0 - 1}{(\beta+1)\lambda_0-\beta}.
\end{equation}
\end{corollary}
This result was also derived in \cite{PiYe86,Ba2004,ZaMo06,DuDoTh09}.
From Proposition~\ref{pro:series_expansion_large} and the proof of Proposition~\ref{prop:lim:throughput} it is seen that $\lambda_0 \rightarrow \infty$ as $\sigma \rightarrow \infty$ and $\beta$ is fixed, and that $\beta (\lambda_0 - 1) \rightarrow \infty$ as $\beta \rightarrow \infty$ and $\sigma$ is fixed. Thus the throughput is approximately $\frac{1}{\beta + 1}$ when either $\sigma$ or $\beta$ is large. This can be understood as follows. For large $\sigma$, the high activity rate allows for configurations close to the maximal independent set: a configuration in which one out of every $\beta+1$ nodes in active. For $\beta$ large, when a node deactivates, a large number of neighboring nodes become eligible for activation. The time until the first such node activates goes to 0 when $\beta$ increases.

\begin{corollary}\label{col:limiting_throughput_beta_small}
Let $\beta \le \eta$. Then
\begin{equation}\label{377777}
\lim_{\sigma \rightarrow \infty} \theta_n(\beta,\eta,\sigma) = 0.
\end{equation}
\end{corollary}
\begin{proof}
From \eqref{eqn:series_expansion_large} it follows that
\begin{equation}\label{eqn:proof_col_limiting_beta_small}
\lambda_0(\sigma) = \sigma^{\frac{1}{1 + \beta}} + \mathcal{O}(1), \quad \sigma \rightarrow \infty.
\end{equation}
Substituting~\eqref{eqn:proof_col_limiting_beta_small} into~\eqref{eqn:asymptotic_throughput}, and using that $f(\beta) > 2 \beta$ when $\beta \le \eta$, yields
\begin{equation}
\theta_n(\beta,\eta,\sigma)= \frac{\sigma (\sigma^{\frac{1}{1 + \beta}} + \mathcal{O}(1))^{\beta-f(\beta)}}{(\beta+1) (\sigma^{\frac{1}{1 + \beta}} + \mathcal{O}(1)) - \beta} \rightarrow 0, \quad (\sigma \rightarrow \infty),
\end{equation}
which gives \eqref{377777}.
\end{proof}

\begin{figure}[h]
 \begin{center}
    \includegraphics[width = 0.6\linewidth]{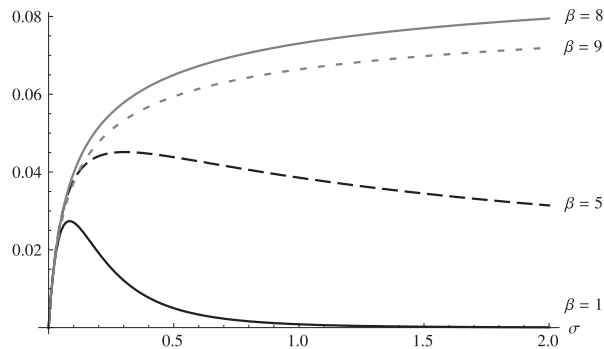}
 \end{center}
  \caption{The throughput $\theta(\beta,\eta,\sigma)$ plotted against $\sigma$ for $\eta = 7$ and various values of $\beta$.}
 \label{fig:limiting_throughput}
\end{figure}

Figure~\ref{fig:limiting_throughput} shows the throughput plotted against the activity rate $\sigma$ for $\eta = 7$ and various values of $\beta$. When $\beta \le \eta$, the throughput gradually drops to 0, whereas for $\beta \ge \eta + 1$, the throughput will eventually converge to the limit $1/(\beta + 1)$. This confirms Corollaries~\ref{col:limiting_throughput_beta_large} and \ref{col:limiting_throughput_beta_small}.

\section{Partition function roots}\label{sec:detailed_study}
In this section we study the roots $\lambda_0,\ldots,\lambda_\beta$ of~\eqref{eqn:equation_lambda} in more detail. In particular, we derive exact infinite-series expressions for the roots that are used in this paper both for numerical purposes (in Section \ref{sec:discussion}) and to prove Corollary~\ref{col:limiting_throughput_beta_small}. These roots are essential in Section~\ref{sec:finite}, where the finite and infinite networks are compared. Our main tool will be the Lagrange inversion theorem (see \cite{DeBruijn81}), and depending on the value of $\sigma$, this gives two different infinite-series expressions. Let $(x)_n = \Gamma(x+n)/\Gamma(x)$ denote the Pochhammer symbol.

\begin{proposition}\label{pro:series_expansion_small}
For small $\sigma > 0$,
\begin{align}
\lambda_0(\sigma) &= 1 + \sum_{l = 1}^\infty \frac{(-1)^{l-1} (\beta l)_{l-1}}{l!} \sigma^l, \label{eqn:series_expansion_small_0}\\
\lambda_j(\sigma) &= \sum_{l = 1}^\infty \frac{(l/\beta)_{l-1}}{l!} w_j^l, \quad j = 1,2,\dots,\beta, \label{eqn:series_expansion_small_j}
\end{align}
where $w_j = \sigma^{1/\beta} {\rm e}^{2 \pi \imath (j - 1/2)/\beta}$ and $\imath = \sqrt{-1}$. The series expansions in~\eqref{eqn:series_expansion_small_0} and~\eqref{eqn:series_expansion_small_j} converge for
\begin{equation}\label{eqn:xi}
0 \le \sigma \le \frac{\beta^\beta}{(\beta+1)^{\beta+1}}=:\xi(\beta),
\end{equation}
and diverge otherwise.
\end{proposition}

\begin{proof}
We first consider the case $j = 0$. Set $\mu_0 = \lambda_0 - 1$, so $\mu_0$ satisfies $\mu_0(1 + \mu_0)^\beta = \sigma$. Hence for small values of $|\sigma|$ we have by Lagrange's inversion theorem
\begin{equation}\label{eqn:proof_series_expansion_small_0}
\mu_0 = \sum_{l = 1}^\infty \frac{1}{l!} \left( \frac{\rm d}{{\rm d}\mu}\right)^{l-1} \left[ \left( \frac{\mu}{\mu(1 + \mu)^{\beta}}\right)^l\right]_{\mu = 0} \sigma^l
= \sum_{l = 1}^\infty \frac{(-1)^{l-1} (\beta l)_{l-1}}{l!} \sigma^l.
\end{equation}
Next we consider the case that $j = 1,\dots,\beta$. We now write \eqref{eqn:equation_lambda} as
\begin{equation}
\lambda^{\beta} (1 - \lambda) = - \sigma, \quad \lambda (1 - \lambda)^{1/\beta} = w_j,
\end{equation}
where
\begin{equation}
w_j = \sigma^{1/\beta} {\rm e}^{2 \pi \imath (j - 1/2)/\beta}.
\end{equation}
Then we get for $|w_j|$ sufficiently small
\begin{equation}\label{eqn:proof_series_expansion_small_j}
\lambda_j = \sum_{l = 1}^\infty \frac{1}{l!} \left( \frac{\rm d}{{\rm d}\lambda}\right)^{l-1} \left[ \left( \frac{\lambda}{\lambda(1 - \lambda)^{1/\beta}}\right)^l\right]_{\lambda = 0} w_j^l
= \sum_{l = 1}^\infty \frac{(l/\beta)_{l-1}}{l!}w_j^l.
\end{equation}

The radii of convergence of the series in~\eqref{eqn:proof_series_expansion_small_0} and~\eqref{eqn:proof_series_expansion_small_j} are easily obtained from the asymptotics
\begin{equation}\label{eqn:asymptotics_gamma_function}
\Gamma(x+1) = x^{x + 1/2} {\rm e}^{-x} \sqrt{2 \pi} (1 + \mathcal{O}(x^{-1}), \quad x \rightarrow \infty,
\end{equation}
of the $\Gamma$-function, used to examine the Pochhammer quantities $(x)_n = \Gamma(x + n)/\Gamma(x)$ and the factorials $l! = \Gamma(l+1)$ that occur in both series. This yields the result that both series converge when $|\sigma| \le \xi(\beta)$ and diverge for $|\sigma| > \xi(\beta)$. When $|\sigma| = \xi(\beta)$ the terms in either series are $\mathcal{O}(l^{-3/2})$.
\end{proof}

\begin{proposition}\label{pro:series_expansion_large}
For large $\sigma > 0$,
\begin{equation}
\lambda_j(\sigma) = \Bigg(\sum_{l = 1}^\infty \frac{\left(\frac{-l}{\beta+1}\right)_{l-1}}{l!} v_j^{-l}\Bigg)^{-1},\quad j = 0,1,\dots,\beta, \label{eqn:series_expansion_large}
\end{equation}
where $v_j = \sigma^{1/(\beta+1)} {\rm e}^{2 \pi \imath j/(\beta+1)}$. The series expansion in~\eqref{eqn:series_expansion_large} converges for
\begin{equation}
\sigma \ge \xi(\beta),
\end{equation}
and diverges otherwise, where $\xi(\beta)$ is given in~\eqref{eqn:xi}.
\end{proposition}

\begin{proof}
We can treat the cases $j = 0$ and $j = 1,\dots,\beta$ simultaneously now. We write \eqref{eqn:equation_lambda} in the form
\begin{equation}
\frac{1}{\lambda} \left(1 - \frac{1}{\lambda}\right)^{\frac{-1}{\beta + 1}} = \left( \frac{1}{\sigma}\right)^{\frac{1}{\beta + 1}} = v^{-1},
\end{equation}
where we let
\begin{equation}\label{eqn:proof_series_expansion_large1}
v^{-1} = v_j^{-1} = \left( \frac{1}{\sigma}\right)^{\frac{1}{\beta + 1}} {\rm e}^{ -2 \pi \imath \frac{j}{ \beta + 1}}, \quad j = 0,1,\dots, \beta
\end{equation}
with $\sigma^{-\frac{1}{\beta + 1}} > 0$ in~\eqref{eqn:proof_series_expansion_large1}. We get for sufficiently large $\sigma$ from Lagrange's inversion theorem (with $u = 1/\lambda$) that
\begin{equation}\label{eqn:proof_series_expansion_large2}
\frac{1}{\lambda_j} = \sum_{l = 1}^\infty \frac{1}{l!} \left( \frac{\rm d}{{\rm d}u}\right)^{l-1} \left[ \left( \frac{u}{u(1-u)^{-1/(\beta + 1)}}\right)^l\right]_{u = 0} v_j^{-l}
= \sum_{l = 1}^\infty \left( \frac{-l}{\beta + 1}\right)_{l-1} \frac{v_j^{-l}}{l!}.
\end{equation}
The Pochhammer quantity $( \frac{-l}{\beta + 1})_{l-1}$ vanishes if and only if $l = 1,2,\dots$ is a multiple of $\beta + 1$.
The radius of convergence of the series in~\eqref{eqn:proof_series_expansion_large2} is again determined by the asymptotics of the $\Gamma$-function in~\eqref{eqn:asymptotics_gamma_function}. Here it must also be used that
\begin{equation}
\Gamma(- J) = \frac{-1}{\Gamma(J + 1)} \frac{\pi}{\sin \pi J},\quad J > 0.
\end{equation}
It follows that the series in~\eqref{eqn:proof_series_expansion_large2} is convergent when $|\sigma| \ge \xi(\beta)$ and divergent when $|\sigma| < \xi(\beta)$. When $|\sigma| = \xi(\beta)$ the terms in the series are $\mathcal{O}(l^{-3/2})$.
\end{proof}

\begin{figure}[h]
 \begin{center}
    \includegraphics[width = 0.6\linewidth]{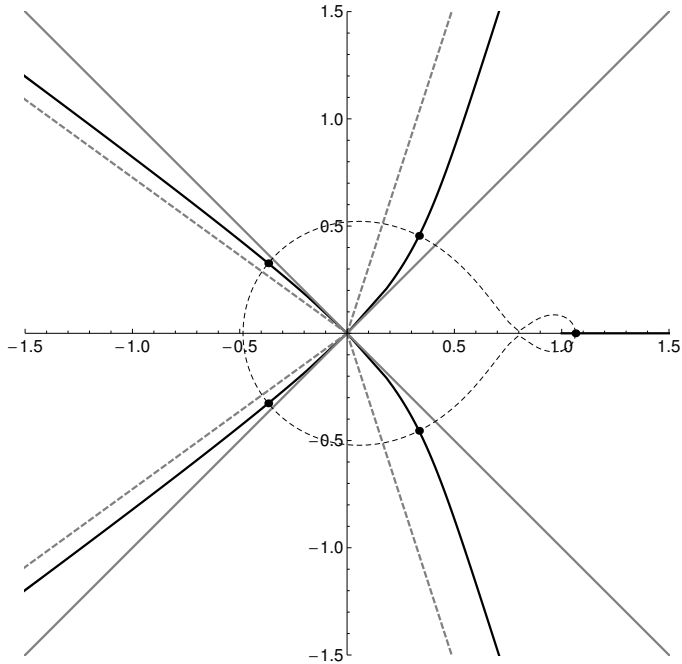}
 \end{center}
  \caption{The roots of~$\lambda^{\beta+1} + \lambda^\beta = \sigma$ as functions of $\sigma$ in~\eqref{eqn:series_expansion_small_0}, \eqref{eqn:series_expansion_small_j} and~\eqref{eqn:series_expansion_large}, for $\beta = 4$.}
 \label{fig:roots_complex_plane}
\end{figure}

Figure~\ref{fig:roots_complex_plane} shows the roots of~\eqref{eqn:equation_lambda} drawn in the complex $\lambda$-plane for $\beta = 4$. Each heavy solid line corresponds to a root as a function of $\sigma$, and the dots represent the threshold $|\sigma|=\xi(\beta)$. The light solid straight line and the dashed straight line illustrate the leading behavior of each root as $\sigma \downarrow 0$ or $\sigma \rightarrow \infty$ according to Propositions~\ref{pro:series_expansion_small} and~\ref{pro:series_expansion_large}, respectively. The dashed curve encircling the origin $0$ and the point $1$ is the image of $v\in\mathbb{C}$ with $|v|=\sigma^{1/(\beta+1)}$, $\sigma = \xi(\beta)$, under the mapping given by the reciprocal of the right-hand side of \eqref{eqn:series_expansion_large} with $v_j$ replaced by $v$.

%

\section{Discussion and outlook}\label{sec:discussion}
The distinguishing feature of this paper is the presence of node interaction when making the tradeoff between hidden nodes and exposed nodes. In order to get a handle on the throughput function (and hence the partition function) we studied the wireless network in the asymptotic regime of infinitely many nodes. This resulted in a tractable limiting expression for the throughput of node zero (and hence of any other node) that allowed us to prove the following two results:

(i) To optimize the throughput, one should always choose a sensing range $\beta$ that is close to the interference range $\eta$, and in fact the optimal sensing range is contained in the interval $[\eta-1,\eta+1]$ (see Proposition \ref{thminterbeta}).

(ii) The sensing range $\beta^*$ that optimizes the throughput equals $\eta-1$ for less aggressive nodes (small $\sigma$) and $\eta+1$ for aggressive nodes (large $\sigma$). In fact, we were able to show the existence of a threshold interval for $\sigma$ that distinguishes these two regimes (Theorem \ref{thm:threshold_interval}). This important result provides (partial) justification for the frequently made assumption that no collisions occur. Indeed, one key take away is that if $\sigma$ is large enough, ruling out all collisions by setting $\beta = \eta+1$ is optimal.

We have further shown that the threshold interval is in many cases small, which implies that one can speak of an almost immediate transition from one regime ($\beta^*=\eta-1$) to the other ($\beta^*=\eta+1$). We have argued that, when the aggressiveness of the nodes is large enough, the system no longer gains from the potential benefits of more flexibility (small $\beta$), and just settles for the situation with no collisions.

We shall now discuss two remaining issues. In Section \ref{sec:finite} we investigate to what extent the asymptotic results give accurate predictions for {\em finite} line networks. In Section \ref{sec:moretop} we investigate whether the notions of two regimes and a critical threshold carry over to more general topologies.

\subsection{Finite versus infinite line networks}\label{sec:finite}
\begin{figure}[h]
 \begin{center}
 \subfigure[$\sigma = 0.25$.]{\label{fig:quality_approximation_theta_vary_beta_1} \includegraphics[width = 0.45\textwidth]{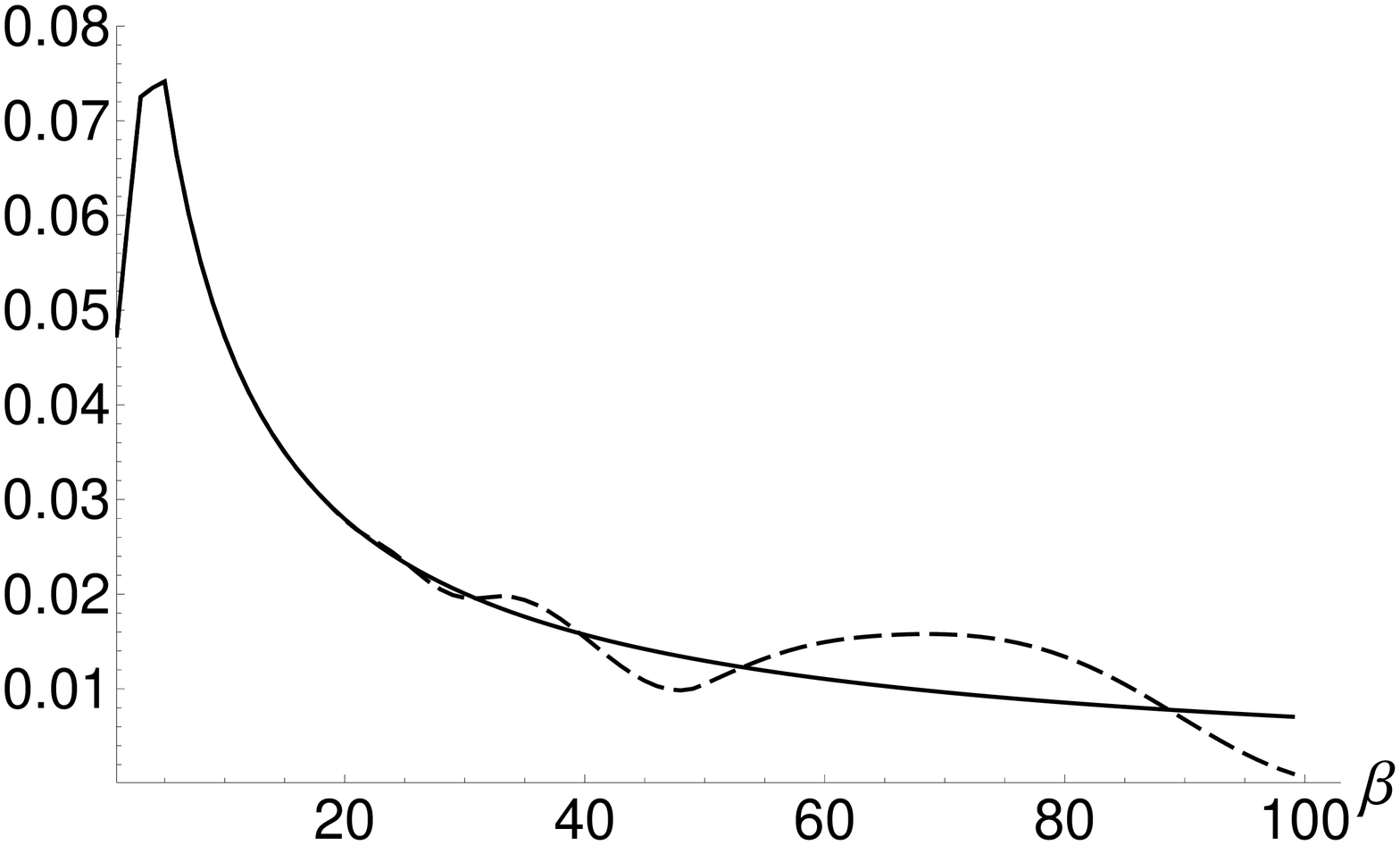}}\hspace{0.2cm}
 \subfigure[$\sigma = 5$.]{\label{fig:quality_approximation_theta_vary_beta_2} \includegraphics[width = 0.45\textwidth]{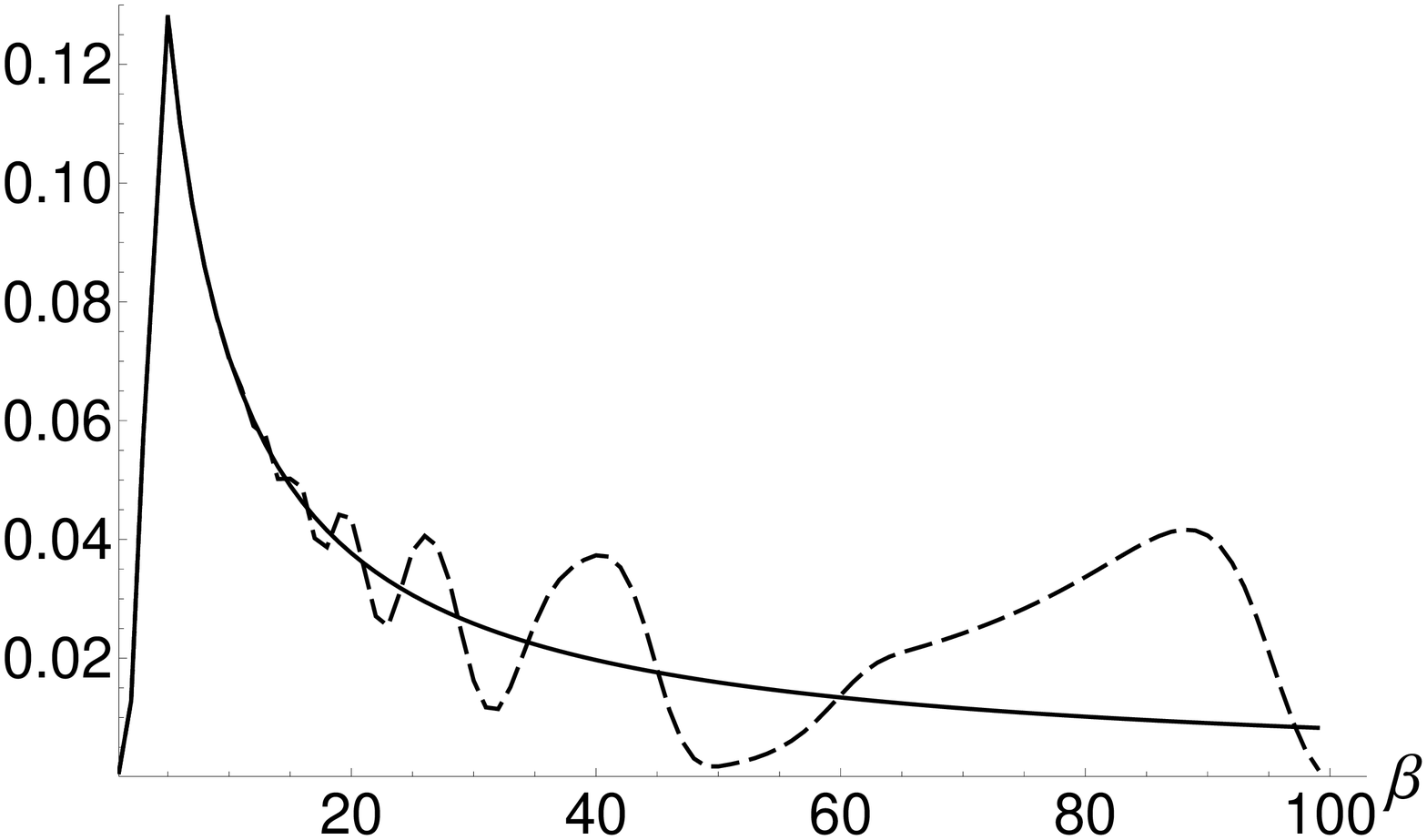}}
 \end{center}
 \caption{The throughput $\theta_n$ (dashed) and $\theta$ (solid) plotted against $\beta$ (with $n=100$).}
 \label{quality_approximation_theta_vary_beta}
\end{figure}

We shall now look at the approximation error $|\theta_n-\theta|$ and the resulting error in the optimal sensing range. To investigate the error we plot $\theta_n$ and $\theta$ in Figure~\ref{quality_approximation_theta_vary_beta}, represented by the dashed line and the solid line, respectively. All results for $\theta_n$ were obtained by using~\eqref{dfg} and~\eqref{eqn:throughput} in combination with the infinite-series expressions for the roots in Section \ref{sec:detailed_study}.

\begin{figure}[h]
 \begin{center}
 \subfigure[$\sigma = 0.25$.]{\label{fig:quality_approximation_theta_vary_n_1} \includegraphics[width = 0.45\textwidth]{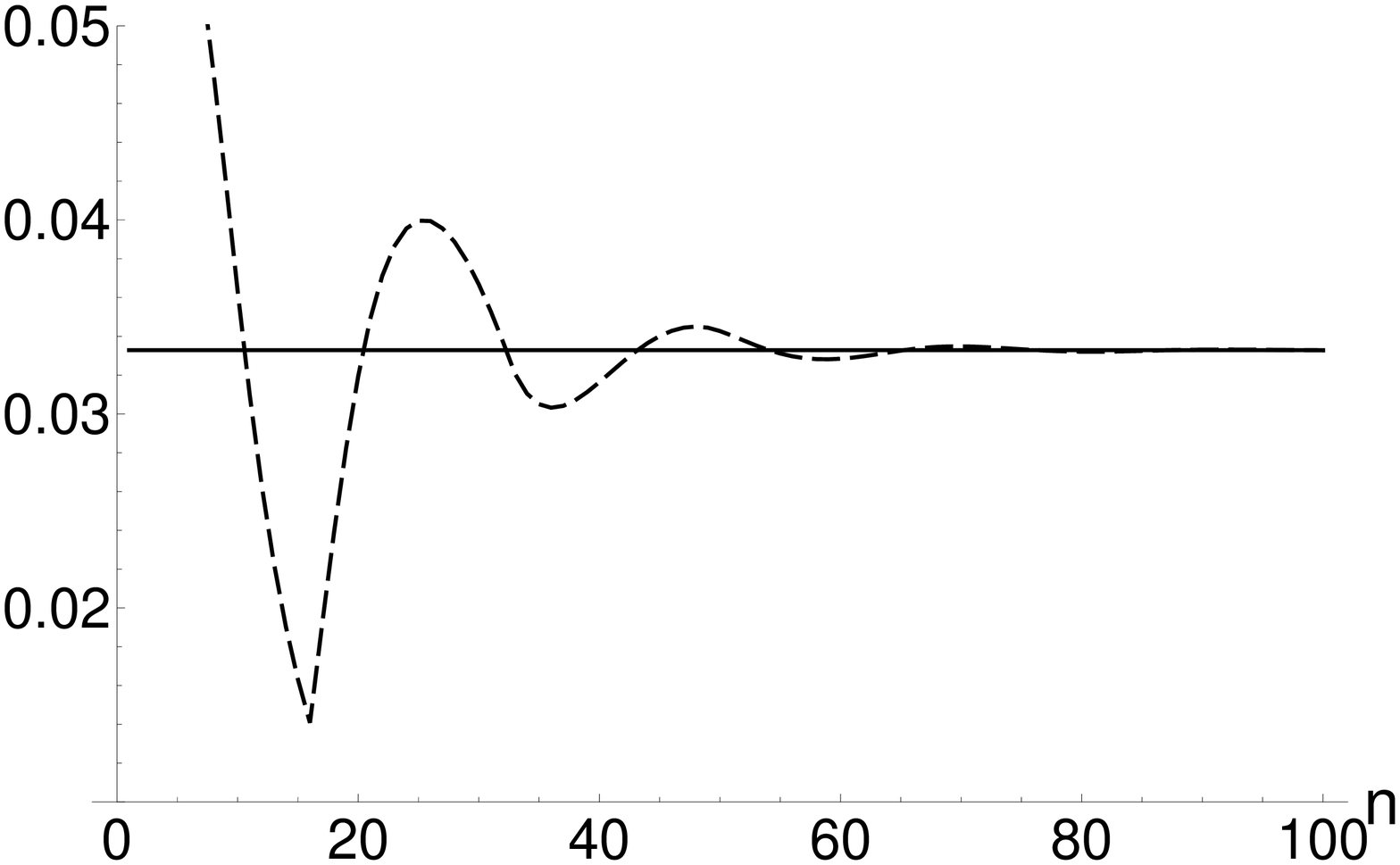}}\hspace{0.2cm}
 \subfigure[$\sigma = 5$.]{\label{fig:quality_approximation_theta_vary_n_2} \includegraphics[width = 0.45\textwidth]{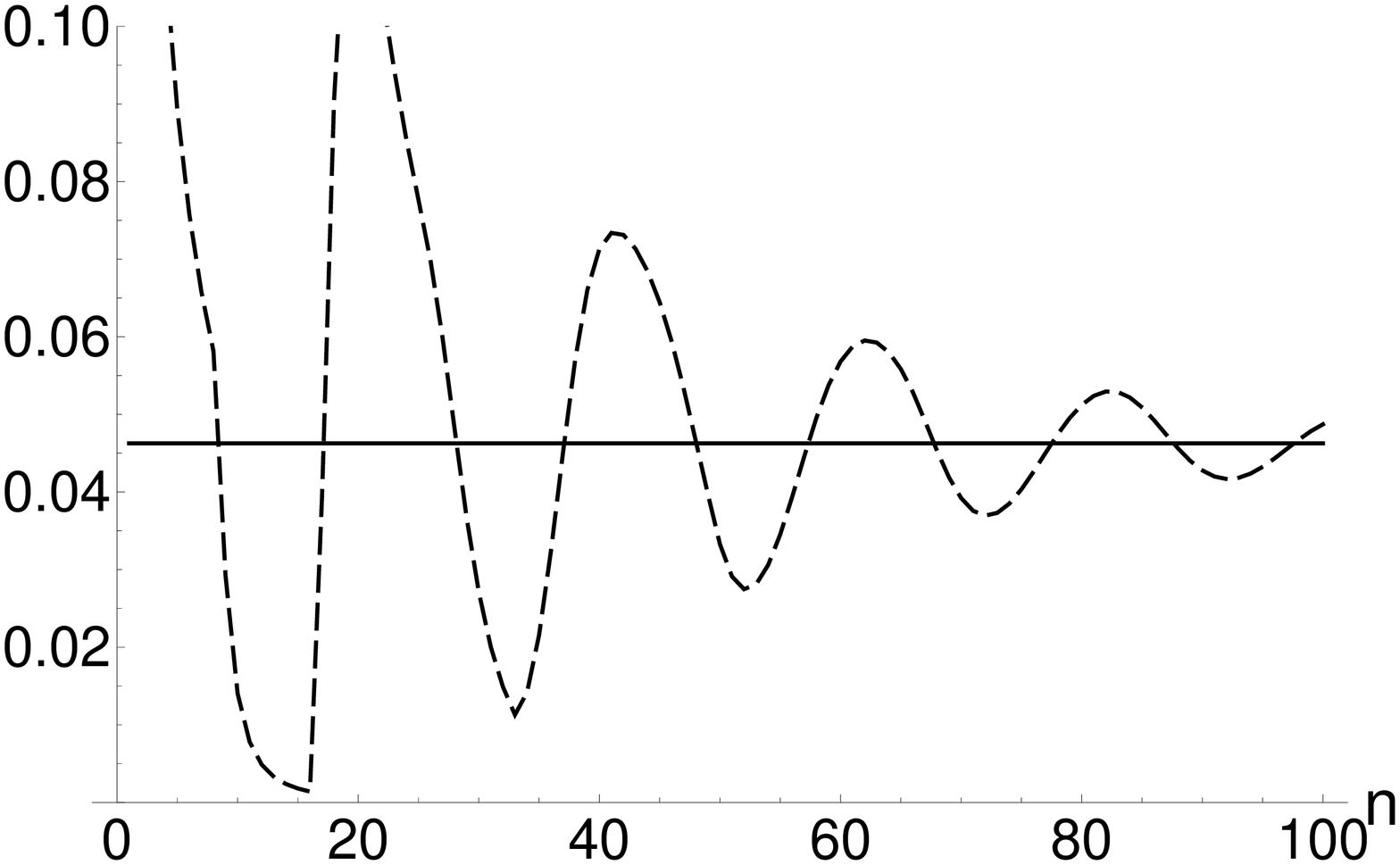}}
 \end{center}
 \caption{The throughput $\theta_n$ (dashed) and $\theta$ (solid) plotted against $n$ (with $\beta=16$).}
 \label{quality_approximation_theta_vary_n}
\end{figure}

 We take $n = 100$ (201 nodes), $\eta = 4$, and we let $\beta$ increase from 1 to 100. In Figure~\ref{fig:quality_approximation_theta_vary_beta_1} $\sigma = 0.25$, and in Figure~\ref{fig:quality_approximation_theta_vary_beta_2} $\sigma = 5$. For $\beta$ small the error $|\theta_n(\beta)-\theta(\beta)|$ is negligible, but the error increases as $\beta$ increases. This can be explained by the observation that for larger $\beta$, the number of roots of~\eqref{eqn:equation_lambda} increases, as does the number of roots discarded by the approximation. This phenomenon becomes more pronounced for larger values of $\sigma$. The non-monotone behavior of $\theta_n$ is caused by the fact that for finite $n$, the system is directed to maximal independent sets of active nodes, in particular for $\sigma$ large, and these sets change dramatically with $\beta$.
The most important observation is that the error $|\theta_n-\theta|$ is small for those values of $\beta$ that lead to a large throughput.
Figure~\ref{quality_approximation_theta_vary_n} is similar to Figure~\ref{quality_approximation_theta_vary_beta}, but instead of fixing~$n$ and varying~$\beta$, we set $\beta = 16$ and vary $n$. In Figure~\ref{fig:quality_approximation_theta_vary_n_1} we take $\sigma = 0.25$ and in Figure~\ref{fig:quality_approximation_theta_vary_n_2} we take $\sigma = 5$. The quality of the approximation increases with $n$.

\begin{figure}[h]
 \begin{center}
 \subfigure[$n = 15$.]{\label{fig:threshold_interval_n15_nolegend} \includegraphics[width = 0.45\textwidth]{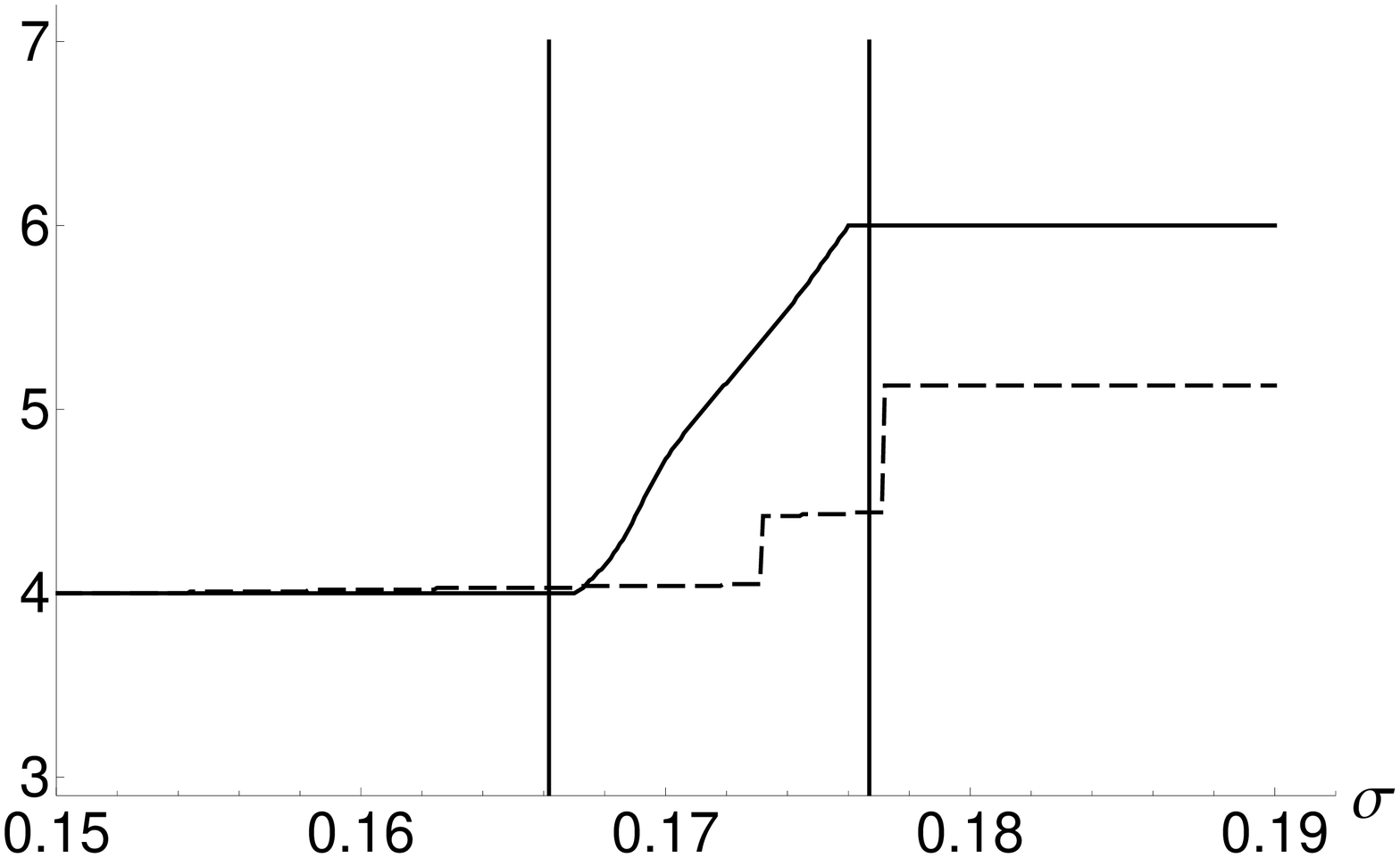}}\hspace{0.2cm}
 \subfigure[$n = 20$.]{\label{fig:threshold_interval_n20_nolegend} \includegraphics[width = 0.45\textwidth]{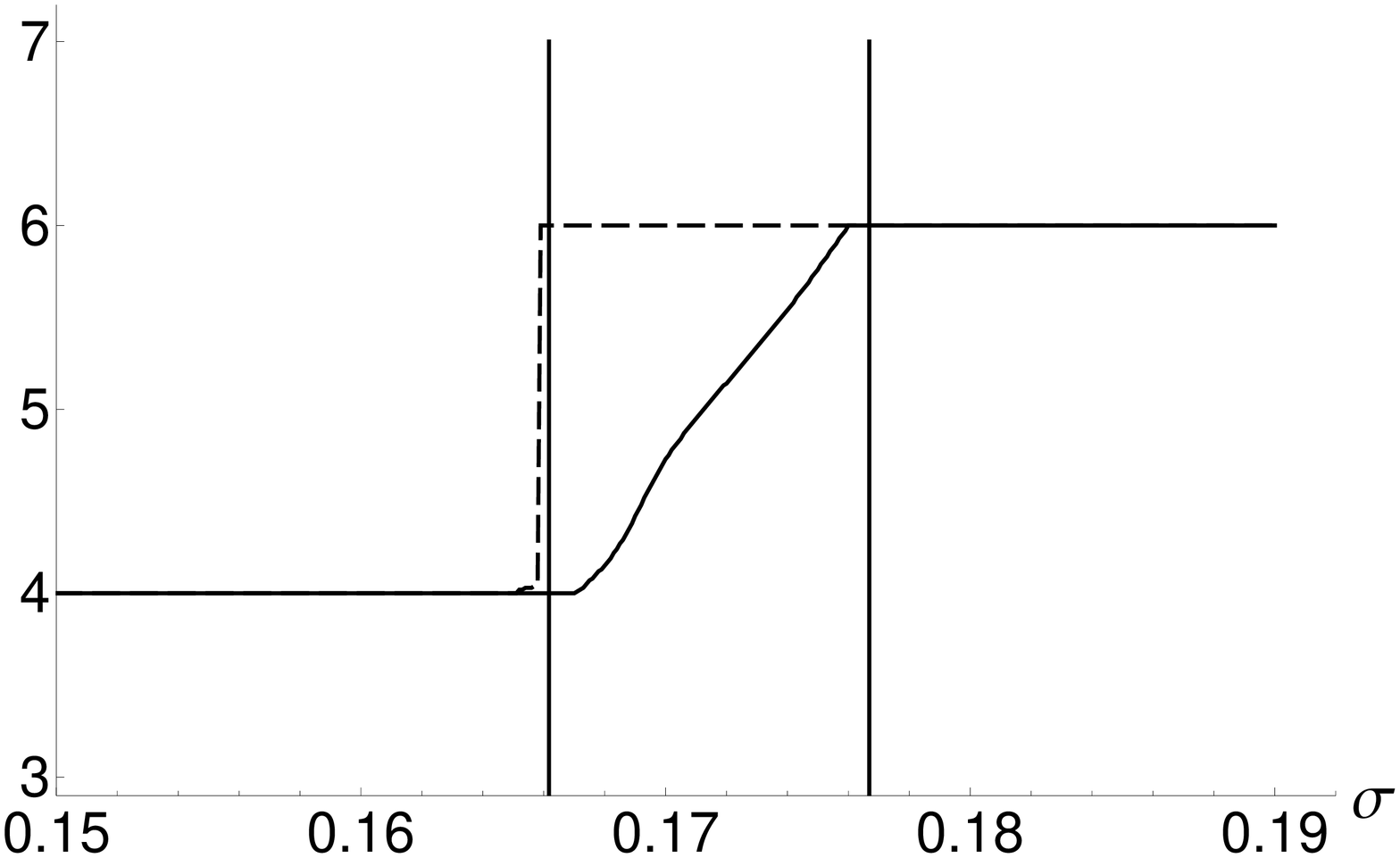}}
 \subfigure[$n = 25$.]{\label{fig:threshold_interval_n25_nolegend} \includegraphics[width = 0.45\textwidth]{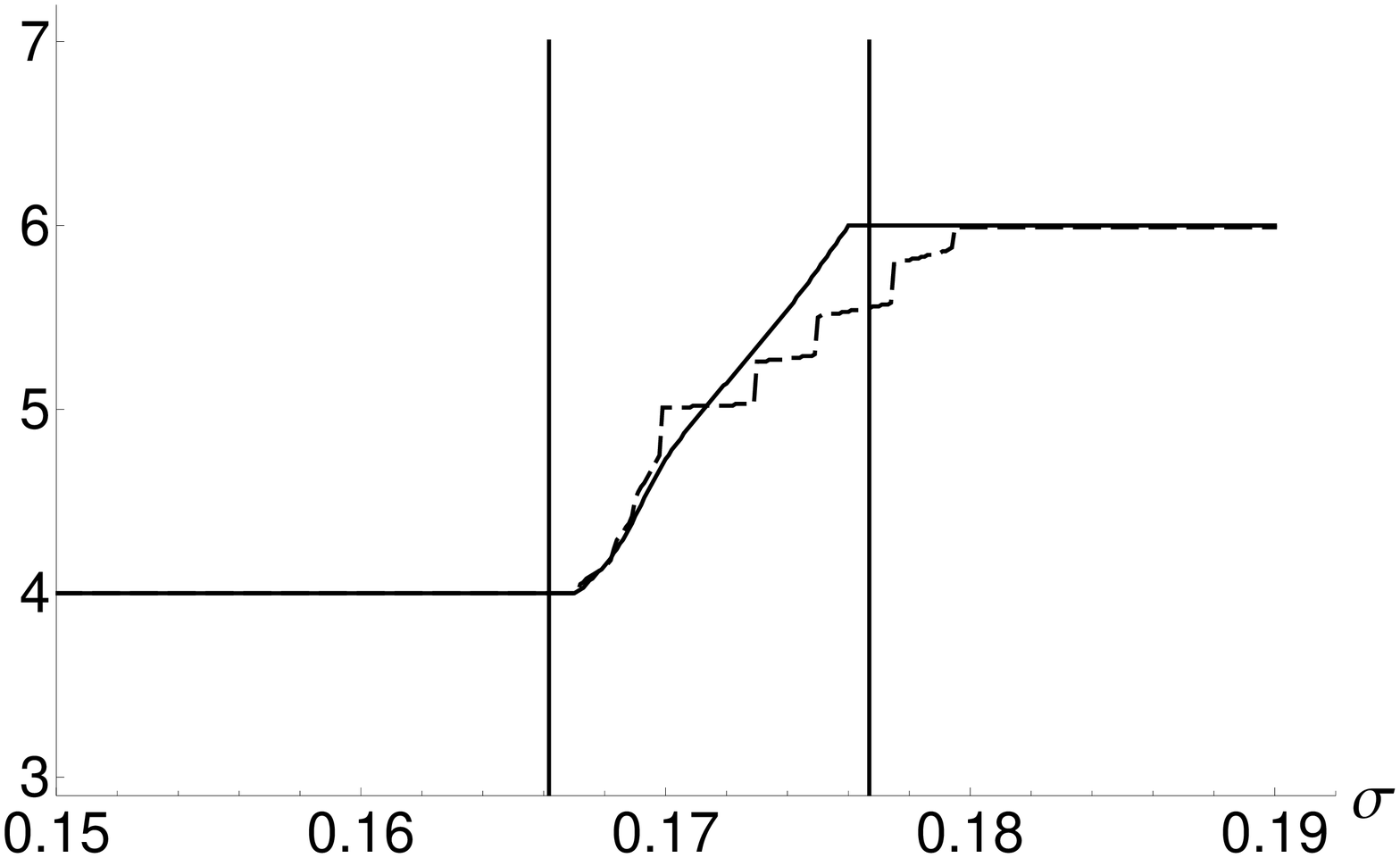}}\hspace{0.2cm}
 \subfigure[$n = 30$.]{\label{fig:threshold_interval_n30_nolegend} \includegraphics[width = 0.45\textwidth]{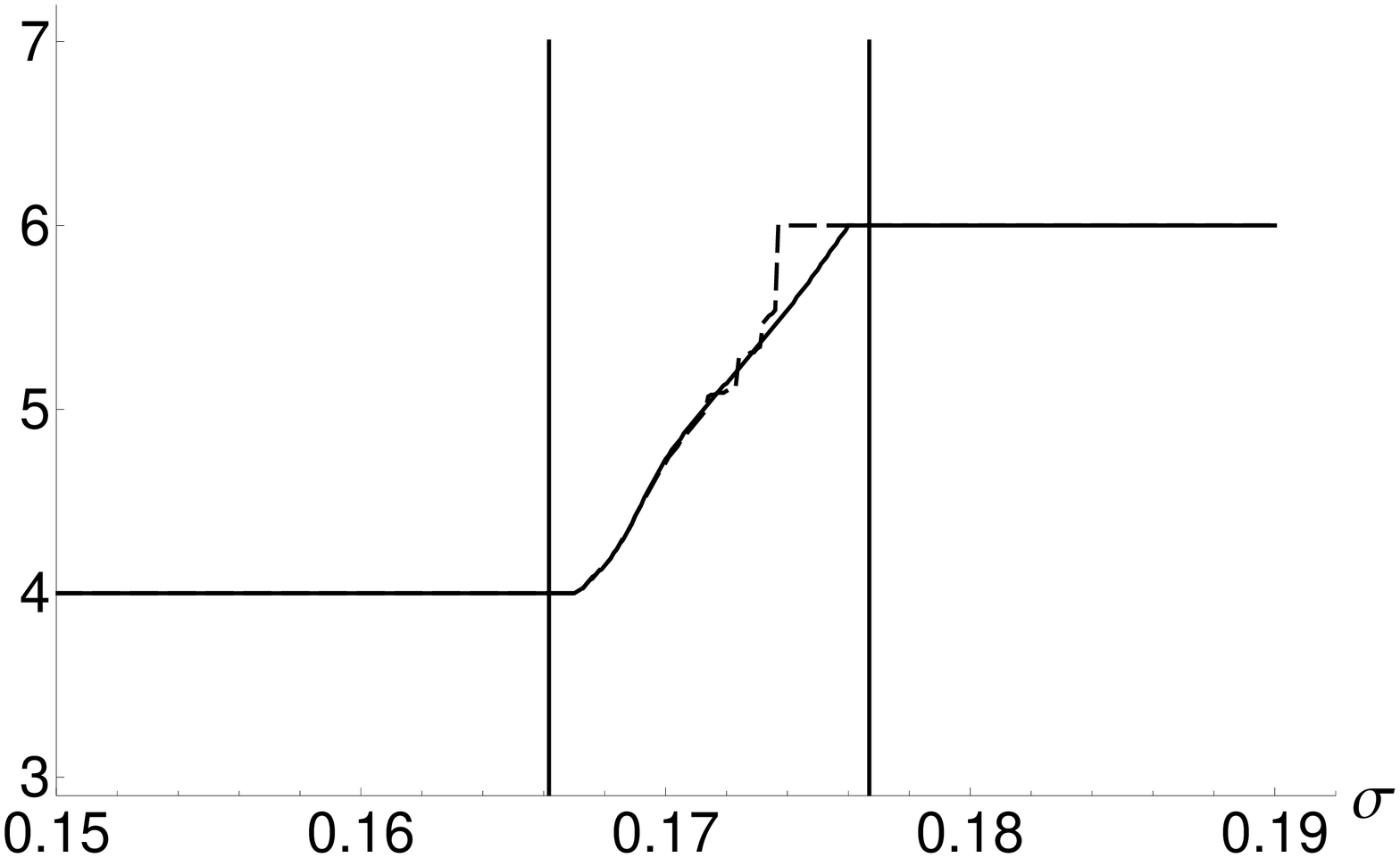}}
 \end{center}
 \caption{The optimal sensing range $\beta^*_n$ (dashed) and $\beta^*$ (solid) plotted against $\sigma$ around the threshold interval for various values of~$n$ and $\eta = 5$.}
 \label{fig:threshold_interval}
\end{figure}

Figure~\ref{fig:threshold_interval} shows the optimal sensing range plotted against $\sigma$, for $\eta = 5$. Each of the Figures~\ref{fig:threshold_interval_n15_nolegend}-\ref{fig:threshold_interval_n30_nolegend} shows the optimal range $\beta_n^*(\sigma)$ for finite $n$. We take $\eta = 5$ for all figures, and let $\sigma$ increase from 0.15 to 0.19. The vertical lines indicate the approximations of the threshold interval from~\eqref{eqn:approximations_sigma_min_max}, and we see that these are sharp. The optimal sensing range~$\beta^*$ for $n \rightarrow \infty$ behaves as predicted by Theorem~\ref{thm:threshold_interval}, jumping from $\eta-1$ before the threshold interval, to $\eta+1$ after this interval, and $\beta^*_n$ shows a similar pattern. We conclude that $n =\infty $ provides a good approximation for the behavior of finite-sized networks, already for small and moderate values of $n$.


\subsection{General topologies}\label{sec:moretop}

%

To investigate more general topologies, we first need a more elaborate description of the model.
In addition to nodes, we introduce directed links between nodes that represent the possibility of transmissions taking place between these nodes. For two nodes to be able to transmit data, we require them to be within (Euclidian) distance~$m$ of each other. We assume links are formed between all nodes within distance $m$. Each node has activation rate~$\sigma$, and the destination of a transmission is chosen uniformly from all links originating from the activating node. The sensing range~$\beta$ and interference range~$\eta$ are also defined using the Euclidian distance.

\begin{figure}[h]
 \begin{center}
 \subfigure[16 nodes on a $4\times 4$ grid.]{\label{fig:grid_topology} \includegraphics[width = 0.3\textwidth]{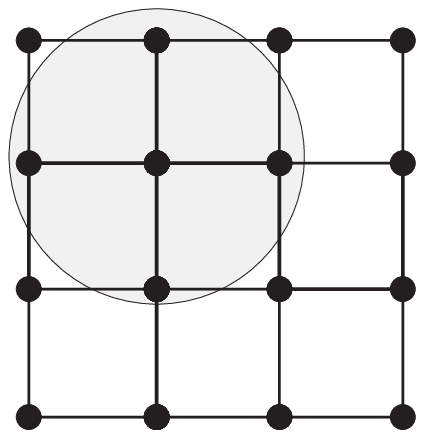}}\hspace{0.3cm}
 \subfigure[The throughput $\theta$ of an arbitrary node in a grid plotted against $\sigma$.]{\label{fig:throughput_wrapped_grid} \includegraphics[width = 0.6\textwidth]{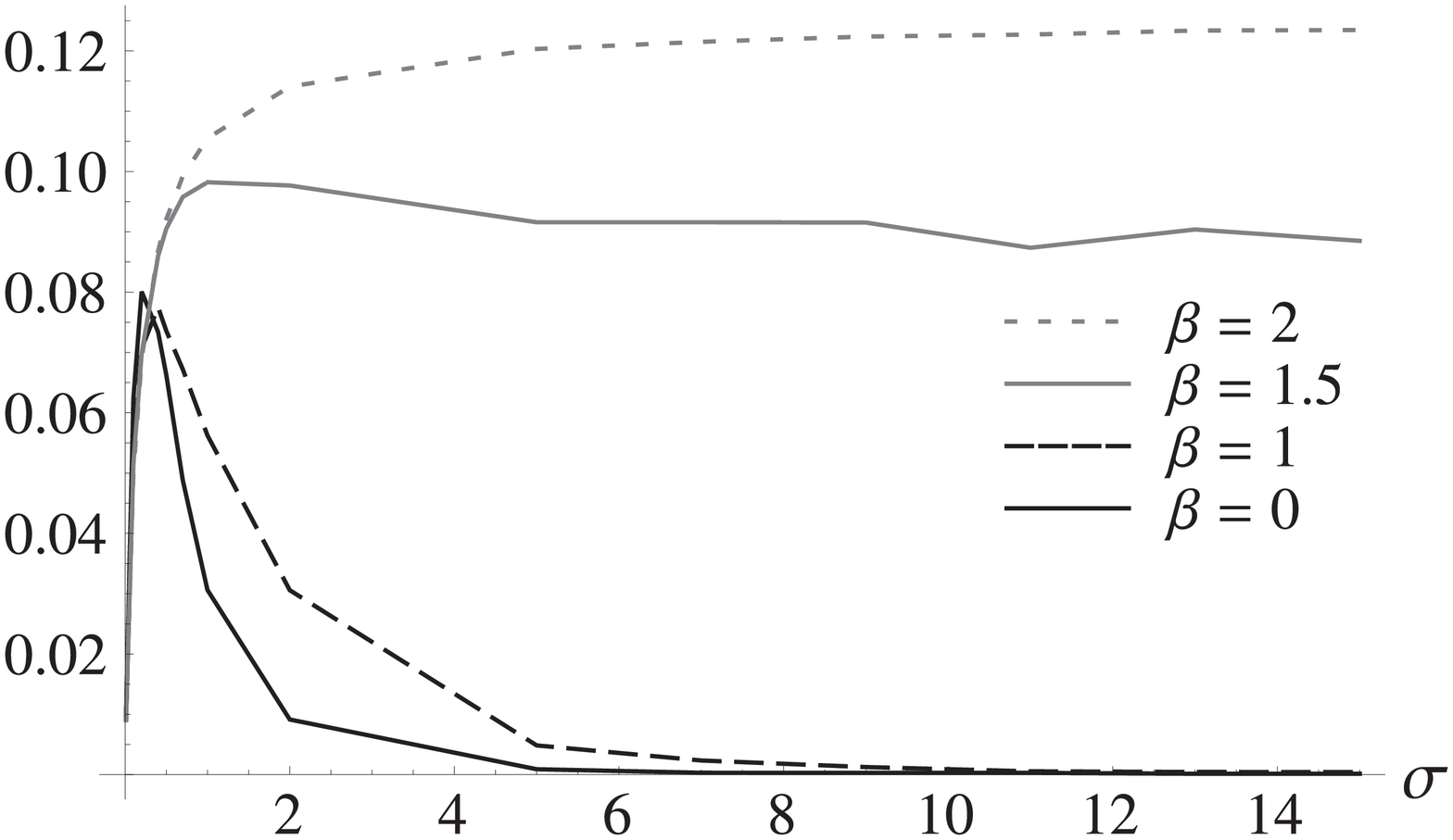}}
 \end{center}
  \caption{A grid network and the corresponding per-node throughput.}
 \label{fig:grid_topology and throughput}
\end{figure}

First we consider $16$ nodes placed on a $4\times 4$ grid at unit distance from each other.
 The grid is wrapped around (top and bottom nodes on any vertical line and left and right nodes on any horizontal line are connected) so that the network is fully symmetric and all nodes have the same environment (and the same throughput), eliminating boundary effects. We set $m = 1$  and construct links between neighboring nodes (see Figure~\ref{fig:grid_topology}). We take $\eta =1$ and $\beta = 0,1,1.5,2$. We run a discrete event simulation of the dynamics described above.

Figure~\ref{fig:throughput_wrapped_grid} shows the average per-node throughput plotted against $\sigma$. For $\sigma$ small we see that $\beta = 0$ (i.e. $\beta = \eta - m$) is throughput-optimal, and for $\sigma$ large it turns out $\beta = 2$ ($\beta = \eta + m$) is optimal. Moreover, when $\beta$ is such that collisions can occur ($\beta < 2$), we see that the throughput decreases when $\sigma$ increases, while for $\beta = 2$ the throughput approaches a non-zero limiting value for large $\sigma$.

We next consider a randomly generated network with 16 nodes. We assume a transmission range of $m=1$ and interference range $\eta = 1.6$. Links are formed between all nodes within distance $m$ and when a node activates, it uniformly chooses a node within distance~$m$ as the receiver.

\begin{figure}[h]
 \begin{center}
    \includegraphics[width = 0.6\linewidth]{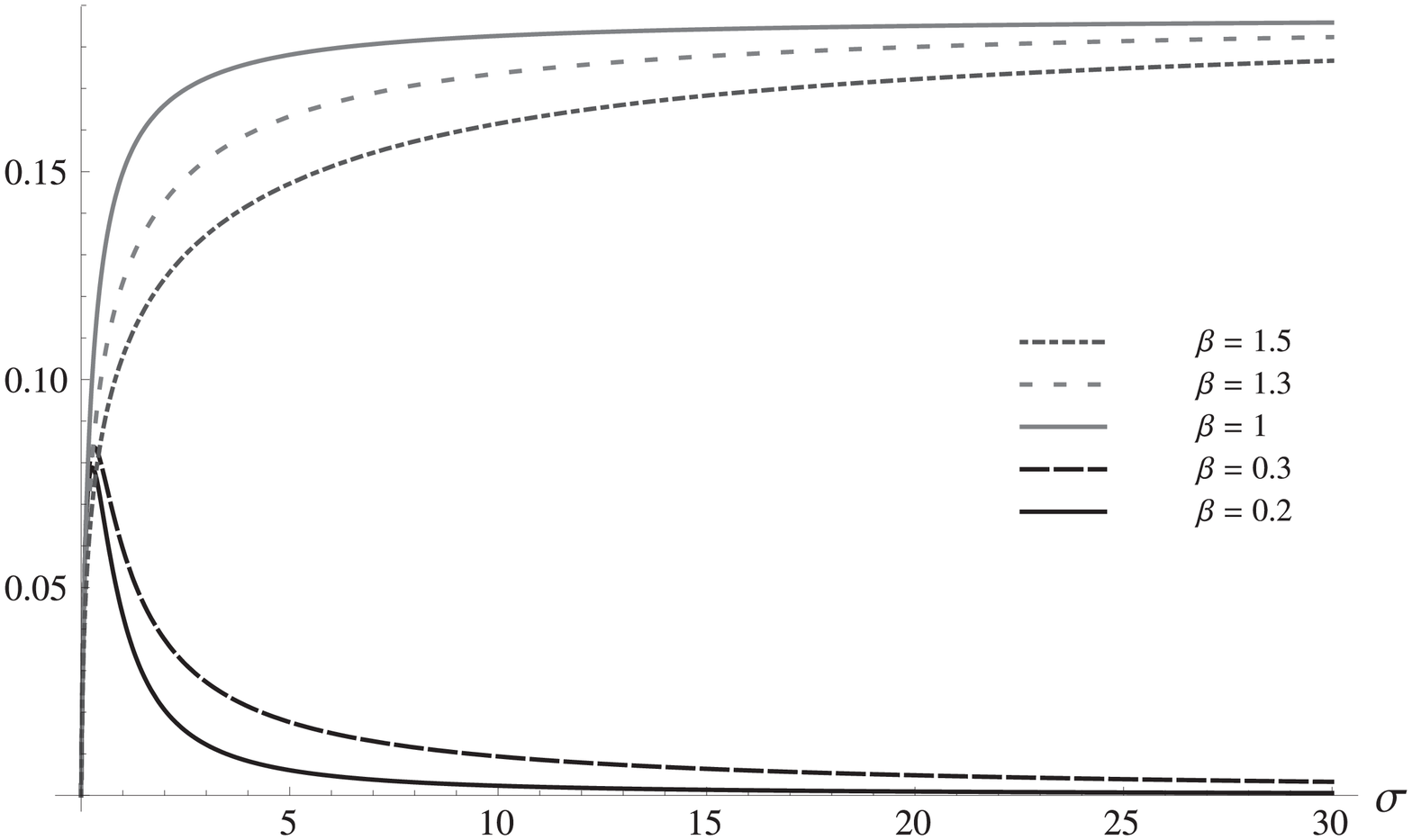}
 \end{center}
  \caption{The average per-node throughput plotted against~$\sigma$.}
 \label{fig:throughput_random}
\end{figure}

The simulation results are shown in Figure~\ref{fig:throughput_random}. The average per-node throughput is plotted against~$\sigma$ for $\beta = 0.2,0.3,1,1.3,1.5$. Figure~\ref{fig:throughput_random} shows resemblance with Figure~\ref{fig:limiting_throughput} for the infinite line. For $\beta$ small the throughput drops as $\sigma$ increases, as a result of collisions. For large $\beta$ collisions are precluded, and the average throughput stabilizes. Moreover, we see that the optimal sensing range $\beta^*$ again depends on $\sigma$. For $\sigma < 0.1$ we have $\beta^* = 0.3$ (this is not visible in the picture), whereas for $\sigma > 0.1$ the optimal sensing range is $\beta^* = 1$.

The tradeoff for individual nodes in an irregular network is more complicated. Although we see a similar threshold interval $(\sigma_{\rm min},\sigma_{\rm max})$ that separates two sensing regimes, the position of the threshold interval and the optimal sensing range may differ between nodes. This depends on the direct surroundings of the node, as well as on the entire network structure.

\subsection{Future work}
Wireless networks equipped with CSMA on complex topologies form highly relevant objects for further study. In particular, we have raised the question whether a threshold interval for the activity rate $\sigma$ exists, which says that the optimal sensing ranges equals $\beta_L$ for $\sigma$ below the interval, and $\beta_U$ for $\sigma$ above the interval. For the two examples in Section \ref{sec:moretop} there is indeed such a threshold interval, but a more thorough study is needed.

Obtaining numerical results for complex topologies with many nodes is challenging. For one thing, the state space no longer decomposes (as with the line network), so that the calculation of the partition function becomes more involved. In determining the stationary distribution, and hence the throughput of nodes, the brute-force method would be to sum over all possible configurations, but that will become computationally cumbersome, already for moderate instances of the network. Alternative approaches would be to use limit theorems, for instance for highly dense networks with many nodes. We conjecture that in such networks we would again find a threshold interval that distinguishes two regimes for the optimal sensing range.

\newpage
\section{Remaining proofs}\label{sec:proofs}

\subsection{Proof of Proposition~\ref{pro:fraction_expansion}}
We write the generating function from~\eqref{eqn:generating_function} as
\begin{equation}
Z(x,\sigma) = \frac{P(x)}{S(x)},
\end{equation}
where
\begin{equation}
P(x) = 1 + \sigma \frac{x^{\beta + 1} - x}{x-1}, \quad S(x) = 1 - x - \sigma x^{\beta + 1}.
\end{equation}
It is shown in~\cite{PiYe86} that the equation $S(x) = 0$ has $\beta + 1$ roots $x_j$, $j = 0,1,\dots,\beta$, and exactly one of them, $x_0$ is real and positive, while $|x_j| > x_0,\ j = 1,\dots,\beta$. To prove Proposition~\ref{pro:fraction_expansion} we first need to establish that these roots are distinct.

\begin{proposition}\label{pro:distinct}
The roots of $S(x) = 0$ are distinct.
\end{proposition}
\begin{proof}
When $S(x) = S'(x) = 0$, we have
\begin{equation}
1 - x - \sigma x^{\beta + 1} = 0 = - 1 - \sigma (\beta + 1) x^{\beta}.
\end{equation}
This implies that $x = 1 + \frac{1}{\beta} > 1$ and so that $\sigma = \frac{1 - x}{x^{\beta + 1}} < 0$. However, $\sigma$ is non-negative.
\end{proof}
Now we proceed with the proof of Proposition~\ref{pro:fraction_expansion}.
Let $\lambda_j = 1/x_j$ so that $\lambda = \lambda_j$ satisfies~\eqref{eqn:equation_lambda}. Using that all zeros of $S$ are distinct, we have for $Z(x,\sigma)$ the partial fraction expansion
\begin{equation}
Z(x,\sigma) = \sum_{j = 0}^\beta \frac{P(x_j)}{S'(x_j)}\frac{1}{x-x_j}.
\end{equation}
Now
\begin{equation}
\frac{P(x_j)}{S'(x_j)} = \frac{1 + \sigma \frac{x_j^{\beta + 1} - x_j}{x_j - 1}}{-1-(\beta + 1) \sigma x_j^{\beta}} = \frac{-x_j^{-\beta}}{1 + (\beta + 1) \sigma x_j^{\beta}} = \frac{-x_j^{-\beta}}{1 + (\beta + 1) \frac{1 - x_j}{x_j}} = \frac{-\lambda_j^{\beta}}{(\beta + 1) \lambda_j - \beta}.
\end{equation}
Here it has been used that
\begin{equation}
\frac{1}{1 - x_j} = \frac{-1}{\sigma x_j^{\beta + 1}}, \quad \sigma x_j^{\beta} = \frac{1 - x_j}{x_j}.
\end{equation}
Then for $|x| < x_0$ we have
\begin{equation}
Z(x,\sigma) = \sum_{j = 0}^{\beta} \frac{P(x_j)}{S'(x_j)} \sum_{i = 0}^\infty \frac{-x^i}{x_j^{i+1}} = \sum_{i=0}^\infty x^i \left( \sum_{j=0}^\beta \frac{\lambda_j^{\beta + 1}}{(\beta + 1) \lambda_j - \beta}\lambda_j^i \right),
\end{equation}
as required.

\subsection{Proof of Proposition \ref{thminterbeta}}

As introduced earlier,
\begin{equation}
\mu_0 = \lambda_0 - 1.
\end{equation}
Then $\mu_0$ depends on $\beta$ and $\sigma$, we have $\mu_0 > 0$, and \
\begin{equation}\label{eqn:relation_mu_0_beta_sigma}
\mu_0(1 + \mu_0)^\beta = \sigma.
\end{equation}
By implicit differentiation with respect to $\beta$, we get from~\eqref{eqn:relation_mu_0_beta_sigma} that
\begin{equation}\label{eqn:dmu_0_dbeta}
\frac{\partial \mu_0}{\partial \beta} = \frac{- \mu_0 (1 + \mu_0) \ln(1 + \mu_0)}{1 + \mu_0 + \beta \mu_0}.
\end{equation}
In particular, both $\mu_0$ and $\lambda_0$ decrease as a function of $\beta > 0$.

Consider the case that $0 \le \beta \le \eta - 1$. Using $\lambda_0^\beta = \frac{\sigma}{\lambda_0 - 1}$ we get
\begin{equation}
\theta(\beta,\eta,\sigma) = \sigma^2 \frac{\lambda_0^{ - 2 \eta}}{(\lambda_0 - 1)((\beta + 1)\lambda_0 - \beta)} = \sigma^2 \frac{\lambda_0^{ - 2 \eta}}{\mu_0(1 + \mu_0 + \beta \mu_0)}.
\end{equation}
Now $\lambda_0^{ - 2 \eta}$ increases as a function of $\beta$, and we shall show that $\mu_0(1 + \mu_0 + \beta \mu_0)$ decreases in $\beta > 0$. We have from~\eqref{eqn:dmu_0_dbeta} that
\begin{align}
\nonumber &\frac{\partial}{\partial \beta}[\mu_0(1 + \mu_0 + \beta \mu_0)]= \frac{\partial}{\partial \beta}[\beta \mu_0^2 + \mu_0 + \mu_0^2] = \mu_0^2 - \frac{1 + 2(1 + \beta)\mu_0}{1 + \mu_0 + \beta \mu_0} \mu_0(1 + \mu_0) \ln(1 + \mu_0)\\
\le{}& \mu_0(\mu_0 - (1 + \mu_0) \ln( 1 + \mu_0)) < 0,
\end{align}
where the last inequality follows from $x \ln x > x-1, x > 1$. We conclude that $\theta$ increases as a function of $\beta \in (0, \eta - 1]$.

Next we consider the case that $\beta \ge \eta + 1$. From $\lambda_0^\beta = \frac{\sigma}{\lambda_0 - 1}$ we get
\begin{equation}
\nonumber \theta(\beta,\eta,\sigma) = \sigma \frac{\lambda_0^\beta}{(\beta + 1) \lambda_0 - \beta} = \frac{\lambda_0 - 1}{(\beta + 1) \lambda_0 - \beta} = \frac{\mu_0}{1 + \mu_0 + \beta \mu_0}.
\end{equation}
Now
\begin{equation}
\frac{\partial}{\partial \beta}\left( \frac{\mu_0}{1 + \mu_0 + \beta \mu_0}\right) = \frac{\frac{\partial \mu_0}{\partial \beta} - \mu_0^2}{(1 + \mu_0 + \beta \mu_0)^2} < 0,
\end{equation}
see~\eqref{eqn:dmu_0_dbeta}, and so $\theta$ decreases as a function of $\beta \ge \eta + 1$. Since $\theta$ depends continuously on $\beta > 0$, the result follows.

\subsection{Proof of Theorem \ref{thm:threshold_interval}}
The proof of the result as stated in Theorem~\ref{thm:threshold_interval} requires expanding several other results. We consider $\beta \in [\eta-1,\eta+1]$ so that
\begin{equation}
\theta(\beta,\eta,\sigma) = \sigma \frac{\lambda_0^{-\eta - 1}}{(\beta + 1) \lambda_0 - \beta} = \sigma \frac{(1 + \mu_0)^{\eta - 1}}{1 + \mu_0 + \beta \mu_0}.
\end{equation}
From~\eqref{eqn:dmu_0_dbeta} it follows from a straightforward but somewhat lengthy computation that
\begin{equation}
 \frac{\partial}{\partial \beta}[\theta(\beta,\eta,\sigma)] = \frac{- \sigma \mu_0 (1 + \mu_0)^{- \eta - 1}}{(1 + \mu_0 + \beta \mu_0)^2} \times \left( 1- (\eta + 2 + \frac{\beta}{1 + \mu_0 + \beta \mu_0}) \ln(1 + \mu_0)\right). \label{eqn:dtheta_dbeta}
\end{equation}
Let
\begin{equation}\label{eqn:F(beta,sigma)}
F(\beta,\sigma) = (\eta + 2 + \frac{\beta}{1 + \mu_0 + \beta \mu_0}) \ln(1 + \mu_0).
\end{equation}
Then we have for $\beta \in [\eta - 1,\eta+1]$ that
\begin{align}
&F(\beta,\sigma) > 1 \Rightarrow \theta {\rm ~increases~strictly~at~} \beta, \label{eqn:F>1}\\
&F(\beta,\sigma) < 1 \Rightarrow \theta {\rm ~decreases~strictly~at~} \beta.\label{eqn:F<1}
\end{align}
We analyze $F(\beta,\sigma)$ in some detail, especially for values of $\beta,\sigma$ such that $F(\beta,\sigma) = 1$. We recall here that $\mu_0 = \mu_0(\beta,\sigma)$ is a function of $\beta$ and $\sigma$ as well.

We fix $\beta > 0$, and we compute
\begin{align}
\nonumber &\frac{\partial}{\partial \beta} F(\beta,\sigma) \\
=&\left[ \frac{\eta + 1}{\mu_0 + 1} + \frac{1 + \beta}{1 + \mu_0 + \beta \mu_0} - \frac{\beta(1 + \beta) \ln(1 + \mu_0)}{(1 + \mu_0 + \beta \mu_0)^2}\right] \frac{\partial \mu_0}{\partial \eta}.
\end{align}
We get from~\eqref{eqn:relation_mu_0_beta_sigma} by implicit differentiation that
\begin{equation}\label{eqn:dmu_0_dsigma}
\frac{\partial \mu_0}{\partial \sigma} = \frac{\mu_0 ( 1+ \mu_0)}{\sigma (1 + \mu_0 + \beta \mu_0)} > 0.
\end{equation}
Furthermore, it is seen from~\eqref{eqn:relation_mu_0_beta_sigma} that $\mu_0(\beta,\sigma) \rightarrow 0$ as $\sigma \downarrow 0$ and that $\mu_0(\beta,\sigma) \rightarrow \infty$ as $\sigma \rightarrow \infty$. Hence, $\mu_0(\beta,\sigma)$ increases from~0 to~$\infty$ as $\sigma$ increases from~0 to~$\infty$. Moreover,
\begin{equation}\label{eqn:inequalities}
\frac{\eta + 1}{\mu_0 + 1} > 0, \quad 1 > \frac{\beta \ln(1 + \mu_0)}{1 + \mu_0 + \beta \mu_0}.
\end{equation}
It follows from~\eqref{eqn:dmu_0_dsigma} and~\eqref{eqn:inequalities} that $\frac{\partial}{\partial \sigma} F(\beta,\sigma) > 0$. Then, from~\eqref{eqn:F(beta,sigma)} and from the fact that $\mu_0$ increases from~0 to~$\infty$ as $\sigma$ increases from~0 to~$\infty$, we have that $F(\beta,\sigma)$ increases from~0 to~$\infty$ as $\sigma$ increases from~0 to~$\infty$. Therefore, for any $\beta > 0$, there is a unique $\sigma = \sigma(\beta)$ such that
\begin{equation}\label{eqn:F=1}
F(\beta,\sigma) = F(\beta,\sigma(\beta)) = 1.
\end{equation}

We shall next show that $\sigma(\beta)$ increases in $\beta \in [\eta- 1,\eta+1]$. By implicit differentiation in~\eqref{eqn:F=1}, we have for $\beta \in [\eta - 1,\eta + 1]$
\begin{equation}\label{eqn:dF_dbeta}
0 = \frac{{\rm d}}{{\rm d} \beta}[F(\beta,\sigma(\beta))] = F_{\beta}(\beta,\sigma(\beta)) + \sigma'(\beta) F_{\sigma}(\beta,\sigma(\beta)),
\end{equation}
where $F_{\beta}$ and $F_{\sigma}$ denote the respective partial derivatives (and $\sigma'(\eta \pm 1)$ is the left and right derivative for $+$ and $-$, respectively). We already know that $F_{\sigma} > 0$, and we shall show now that $F_{\beta}(\beta,\sigma(\beta)) < 0$. To that end, we compute, using definition~\eqref{eqn:F(beta,sigma)} of $F$ and~\eqref{eqn:dmu_0_dbeta} that
\begin{equation}
\frac{\partial}{\partial \beta}[F(\beta,\sigma)] = - \ln(1 + \mu_0) \Big[ (\eta + 2 + \frac{\beta}{1 + \mu_0 + \beta \mu_0}) \frac{\mu_0}{1 + \mu_0 + \beta \mu_0} - \frac{1 + \mu_0 - \beta (1 + \beta) \frac{\partial \mu_0}{\partial \beta}}{(1 + \mu_0+ \beta \mu_0)^2}\Big].
\end{equation}
Next, from~\eqref{eqn:F(beta,sigma)} and~\eqref{eqn:F=1} we have that
\begin{equation}
\mu_0 \ge \ln(1 + \mu_0) = \frac{1}{\eta + 2 + \frac{\beta}{1 + \mu_0 + \beta \mu_0}},
\end{equation}
and so
\begin{align}
\nonumber &\frac{\partial F}{\partial \beta}(\beta,\sigma(\beta)) \le - \ln(1 + \mu_0) \left[ \frac{1}{1 + \mu_0 + \beta \mu_0} - \frac{1 + \mu_0 - \beta(1 + \beta) \frac{\partial \mu_0}{\partial \beta}}{(1 + \mu_0 + \beta \mu_0)^2}\right]_{\sigma = \sigma(\beta)}\\
\nonumber =& \frac{- \beta \ln(1 + \mu_0)}{(1 + \mu_0 + \beta \mu_0)^2}\left[\mu_0 + (1 + \beta) \frac{\partial \mu_0}{\partial \beta} \right]_{\sigma = \sigma(\beta)}\\
=& \frac{-\mu_0 \beta \ln(1 + \mu_0)}{(1 + \mu_0 + \beta \mu_0)^2}\left[ 1 - (1 + \beta) \frac{(1 + \mu_0) \ln(1 + \mu_0)}{1 + \mu_0 + \beta \mu_0}\right]_{\sigma  =\sigma(\beta)},
\end{align}
where~\eqref{eqn:dmu_0_dbeta} has been used once more. Finally, from~\eqref{eqn:F(beta,sigma)} and~\eqref{eqn:F=1},
\begin{equation}
(1 + \beta) \frac{(1 + \mu_0) \ln(1 + \mu_0)}{1 + \mu_0 + \beta \mu_0}\Big|_{\sigma = \sigma(\beta)} = \frac{(1 + \beta)(1 + \mu_0)}{(\eta + 2)(1 + \mu_0 + \beta \mu_0) + \beta} \Big|_{\sigma = \sigma(\beta)} < 1,
\end{equation}
since $0 < \beta \le \eta + 1$ and $\mu_0 > 0$. Hence, $F_{\beta}(\beta,\sigma(\beta)) < 0$ as required. It now follows from~\eqref{eqn:dF_dbeta} and from $F_{\sigma}(\beta,\sigma(\beta)) > 0$ that $\sigma'(\beta) > 0$ when $\beta \in [\eta - 1,\eta + 1]$.

We have now shown that $\sigma(\beta)$ increases in $\beta \in [\eta - 1,\eta+1]$. Next we let
\begin{equation}
\sigma_{\rm min} := \sigma (\eta - 1) < \sigma (\eta + 1) =: \sigma_{\rm max}.
\end{equation}
For $\sigma \in [\sigma_{\rm min},\sigma_{\rm max}]$ there is defined the inverse function $\beta(\sigma) \in [\eta - 1,\eta + 1]$ that increases in $\sigma$. It follows then from
\begin{equation}
F(\beta(\sigma),\sigma) = 1, \quad F_{\beta}(\beta(\sigma),\sigma) < 0
\end{equation}
and~\eqref{eqn:dtheta_dbeta}-\eqref{eqn:F<1} that $\theta(\beta,\eta,\sigma)$ is maximal at $\beta = \beta(\sigma)$ when $\sigma \in [\sigma_{\rm min},\sigma_{\rm max}]$.

We shall now complete the proof of Theorem~\ref{thm:threshold_interval}. Let $\beta \in [\sigma_{\rm min},\sigma_{\rm max}]$, and assume that $\sigma \le \sigma_{\rm min}$. Then $\sigma < \sigma(\beta)$ and so $F(\beta,\sigma) < F(\beta,\sigma(\beta)) = 1$ since $F$ increases in $\sigma$. Hence, $\theta$ strictly decreases at $\beta$. Similarly, $\theta$ strictly increases at $\beta \in (\eta - 1,\eta + 1)$ when $\sigma \ge \sigma_{\rm max}$. It follows that $\theta$ strictly decreases in $\beta \in [\eta - 1,\eta + 1]$ when $\sigma \le \sigma_{\rm min}$ and that $\theta$ strictly increases in $\beta \in [\eta - 1,\eta + 1]$ when $\sigma \ge \sigma_{\rm max}$. Finally, when $\sigma \in (\sigma_{\rm min},\sigma_{\rm max})$, we have that
\begin{equation}
F(\eta - 1,\sigma) > F(\eta - 1,\sigma_{\rm min}) = 1 = F(\eta + 1,\sigma_{\rm max}) > F(\eta + 1,\sigma),
\end{equation}
showing that $\theta$ strictly increases at $\beta = \eta - 1$ and strictly decreases at $\beta = \eta + 1$, and assumes its maximum at $\beta = \beta(\sigma)$.

\subsection{Proof of Theorem \ref{thm:length_threshold_interval}}
We shall show below that
\begin{equation}\label{eqn:proof_length_ci_ineq}
(\eta+2 + \frac{\eta - 1}{1 + \eta \kappa}) \ln(1 + \kappa) < 1 < (\eta + 2 + \frac{\eta + 1}{1 + (\eta + 2) \kappa} )\ln(1 + \kappa)
\end{equation}
where $\kappa = \tau/(\eta+1)$. Assuming this, we recall that (for fixed $\beta > 0$) $\mu_0$ strictly increases in $\sigma$ and vice versa. When now
\begin{equation}
\sigma_{-} = \kappa(1 + \kappa)^{\eta - 1},
\end{equation}
then $\kappa = \mu_0 (\beta = \eta - 1,\sigma_-)$ and we have that $F(\eta - 1,\sigma_{-}) < 1$. So $\sigma_{-} < \sigma_{\rm min}$ since $F$ is increasing in $\sigma$. Similarly, when
\begin{equation}
\sigma_{+} = \kappa(1 + \kappa)^{\eta+1},
\end{equation}
we have that $\kappa = \mu_0(\beta = \eta+1,\sigma_+)$ and then from~\eqref{eqn:proof_length_ci_ineq} that $F(\eta + 1,\sigma_{+}) > 1$ and so $\sigma_{+} > \sigma_{\rm max}$. Therefore,
\begin{align}
\nonumber \sigma_{\rm max} - \sigma_{\rm min} &< \sigma_{+} - \sigma_{-} = \kappa (1 + \kappa)^{\eta + 1} ((1 +\kappa)^2 - 1) = 2 \left( 1 + \frac{\tau}{\eta + 1}\right)^{\eta -1} \left( \frac{\tau}{\eta + 1}\right)\left( 1 + \frac{\tau}{\eta + 1}\right)\\
 &\le 2 {\rm e}^{\tau} \left( \frac{\tau}{\eta + 1}\right)^2 (1 + \frac{\tau}{\eta + 1}).
\end{align}

This proves Theorem~\ref{thm:length_threshold_interval}(i). It remains to show~\eqref{eqn:proof_length_ci_ineq}. As to the first inequality in~\eqref{eqn:proof_length_ci_ineq} we have
\begin{align}
\nonumber &1 - (\eta + 2 + \frac{\eta - 1}{1 + \eta \kappa}) \ln(1 + \kappa) > 1 - (\eta + 2+ \frac{\eta - 1}{1 + \eta \kappa}) \kappa\\
={}& \frac{1}{1 + \eta \kappa}(1 - (\eta +1) \kappa - \eta (\eta + 2) \kappa^2) > \frac{1}{1 + \eta \kappa} (1 - (\eta + 1) \kappa - ((\eta + 1) \kappa)^2) = 0
\end{align}
since $1 - \tau - \tau^2 = 0$ and $(\eta + 1)\kappa = \tau$. As to the second inequality of~\eqref{eqn:proof_length_ci_ineq} we have
\begin{align}
\nonumber & 1 - (\eta + 2 + \frac{\eta + 1}{1 + (\eta + 2) \kappa}) \ln(1 + \kappa) < 1 - (\eta + 2 + \frac{\eta + 1}{1 + (\eta + 2) \kappa})(\kappa - \frac{1}{2} \kappa^2)\\
={}& \frac{1}{1 + (\eta + 2) \kappa}\Big( 1 - (\eta + 1) \kappa - ((\eta + 1)\kappa)^2 -  \kappa^2 (\eta + 3/2 - \frac{1}{2} (\eta + 2)^2 \kappa)\Big).
\end{align}
As before
\begin{equation}
1 - (\eta + 1) \kappa - ((\eta + 1) \kappa)^2 = 0
\end{equation}
and
\begin{equation}
\eta + \frac{3}{2} - \frac{1}{2} (\eta + 2)^2 \kappa = \eta + \frac{3}{2} - \frac{(\eta + 2)^2}{2 (\eta + 1)} \tau > 0, \quad \eta \ge 0
\end{equation}
since $\tau = \frac{1}{2}(\sqrt{5} - 1) < \frac{3}{4}$ (which is the minimum value of $2(\eta + 3/2)(\eta + 1)(\eta+2)^{-2}$ for $\eta \ge 0)$. This shows the second inequality in~\eqref{eqn:proof_length_ci_ineq}.

We next prove Theorem~\ref{thm:length_threshold_interval}(ii), and for this we need the following result:
\begin{proposition}\label{pro:approximations_sigma_beta}
With $\beta = \eta+ \gamma$ where $-1 \le \gamma \le 1$, we have
\begin{equation}\label{eqn:approximation_sigmas}
\sigma(\beta) = \mu(1 + \mu)^{\eta + \gamma},
\end{equation}
where
\begin{equation}\label{eqn:equation_p11}
\mu = \frac{\tau}{\eta + \alpha + \mathcal{O}(\eta^{-1})}, \quad \alpha = \frac{(5 + 2 \gamma)\tau + 1}{2 (2 \tau + 1)},
\end{equation}
and the $\mathcal{O}$ holds uniformly in $\gamma \in [-1,1]$.
\end{proposition}

\begin{proof}
We have $\sigma(\beta) = \mu(1 + \mu)^{\beta}$ where $\mu$ is the unique solution of the equation
\begin{equation}\label{eqn:equation_mu}
(\eta + 2 + \frac{\beta}{1 + (1 + \beta) \mu}) \ln(1 + \mu) = 1.
\end{equation}
We know from the proof of Theorem~\ref{thm:length_threshold_interval}(i) that $\mu = \mathcal{O}(\eta^{-1})$. Multiplying~\eqref{eqn:equation_mu} by $1 +(1 + \beta) \mu$ and developing
\begin{equation}
\ln (1 + \mu) = \mu - \frac{1}{2} \mu^2 + \mathcal{O}(\mu^3),
\end{equation}
we get
\begin{equation}
(\eta \beta + \frac{1}{2} \eta + \frac{3}{2} \beta + 1) \mu^2 + (\eta + 1) \mu - 1 = \frac{1}{2} (\eta + 2)(\beta + 1) \mu^3 + \mathcal{O}(\eta^{-2}).
\end{equation}
Next let $\alpha \in \mathds{R}$ be independent of $\eta$ and use $\beta = \eta + \gamma$ to write
\begin{equation}
\eta \beta + \frac{1}{2} \eta + \frac{3}{2} \beta + 1 = (\eta + \alpha)^2 + (2 + \gamma - 2 \alpha) \eta + \frac{3}{2} \gamma + 1 - \alpha^2.
\end{equation}
Together with $\eta + 1 = \eta + \alpha + 1 - \alpha$, we obtain
\begin{align}
\nonumber & (\eta + \alpha)^2 \mu^2 + (\eta + \alpha) \mu - 1\\
={}& \frac{1}{2} (\eta + 2)(\eta + \gamma + 1) \mu^3 - ((2 + \gamma - 2 \alpha) \eta + \frac{3}{2} \gamma + 1 - \alpha^2 ) \mu^2 - (1 - \alpha) \mu + \mathcal{O}(\eta^{-2}). \label{eqn:proof_p11}
\end{align}
We now take $\alpha$ such that the whole second member of~\eqref{eqn:proof_p11} is $\mathcal{O}(\eta^{-2})$. Using that $\mu =  \frac{\tau}{\eta} + \mathcal{O}(\eta^{-2})$, this leads to
\begin{equation}
\frac{1}{2} \tau^3 - (2 + \gamma - 2 \alpha) \tau^2 - (1 - \alpha) \tau = 0,
\end{equation}
and this yields the $\alpha$ in~\eqref{eqn:equation_p11}. The polynomial $x^2 + x - 1 = 0$ has a zero of first order at $x = \tau$. Hence with $\alpha$ as in~\eqref{eqn:equation_p11} we see from $(\eta + \alpha)^2 \mu^2 + (\eta + \alpha) \mu - 1 = \mathcal{O}(\eta^{-2})$ that $(\eta + \alpha) \mu = \tau + \mathcal{O}(\eta^{-2})$. This gives the result.
\end{proof}

Now we proceed to prove Theorem~\ref{thm:length_threshold_interval}(ii). We use the result of Proposition~\ref{pro:approximations_sigma_beta}. Thus
\begin{align}
\sigma(\eta + \gamma) &= \mu(1 + \mu)^{\eta + \gamma}, \label{eqn:approximation_sigma_min}\\
\mu &= \frac{\tau}{\eta + \alpha + \mathcal{O}(\eta^{-1})} = \frac{\tau}{\eta + \alpha}(1 + \mathcal{O}(\eta^{-2})). \label{eqn:approximation_sigma_max}
\end{align}
By elementary considerations
\begin{align}
\nonumber \sigma(\eta +& \gamma) = \frac{\tau}{\eta + \alpha}(1 + \frac{\tau}{\eta + \alpha})^{\eta + \gamma}(1 + \mathcal{O}(\eta^{-2}))\\
\nonumber ={}& \frac{\tau}{\eta + \alpha} \exp[(\eta + \gamma)(\frac{\tau}{\eta + \alpha} - \frac{\tau^2}{2(\eta + \alpha)})](1 + \mathcal{O}(\eta^{-2}))\\
={}& \frac{\tau {\rm e}^{\tau}}{\eta + \alpha}(1 + \frac{(\gamma - \alpha)\tau - \frac{1}{2}\tau^2}{\eta})(1 + \mathcal{O}(\eta^{-2})).
\end{align}
Then letting $\gamma = \pm 1$ and
\begin{equation}\label{eqn:alpha}
\alpha(1) =  \frac{7 \tau + 1}{2(2 \tau + 1)}, \quad \alpha(-1) = \frac{3 \tau + 1}{2(2 \tau + 1)}
\end{equation}
in accordance with Proposition~\ref{pro:approximations_sigma_beta}, it follows that
\begin{align}
\nonumber \sigma(\eta + 1) - \sigma(\eta - 1) ={}& \frac{\tau {\rm e}^{\tau}}{\eta^2}\Big(\alpha(-1) - \alpha(1) + (1 - \alpha(1))\tau + (1 + \alpha(-1))\tau \Big) + \mathcal{O}(\eta^{-3})\\
={}& \frac{\tau {\rm e}^\tau}{\eta^2} \frac{2 \tau^2}{2 \tau + 1} + \mathcal{O}(\eta^{-3}).
\end{align}
Finally, it follows easily from $\tau^2 + \tau = 1$ that $\tau^3(7 + 4\tau) = 2\tau + 1$.

\subsection{Proof of Proposition \ref{prop:lim:throughput}}
Since $\sigma > 0$ is fixed, it follows from (see the proof of Theorem~\ref{thm:length_threshold_interval})
\begin{equation}\label{eqn:equation_speed_crit_int1}
\sigma_{\rm max} < \sigma_{+} = \frac{\tau}{\eta + 1} \left(1 + \frac{\tau}{\eta +1}\right)^{\eta + 1} < \frac{\tau {\rm e}^{\tau}}{\eta + 1}
\end{equation}
that $\sigma_{\rm max} < \sigma$ when $\eta$ is large enough. Then by Theorem~\ref{thm:threshold_interval}
\begin{align}
\max \theta &= \theta(\eta + 1) = \frac{\lambda_0 - 1}{(\eta + 2) \lambda_0 - \eta - 1} = \frac{\mu_0}{(\eta + 2) \mu_0 + 1} = \frac{1}{\eta + 2} \frac{1}{1 + \frac{1}{(\eta + 2) \mu_0}},
\end{align}
where $\mu_0$ is the unique positive real $\mu$ root of $\mu (1 + \mu)^{\eta + 1} = \sigma$. We shall show that
\begin{align}\label{eqn:equation_speed_crit_int2}
(\eta + 2) \mu_0 &\ge \ln \sigma,\\
\label{eqn:equation_speed_crit_int3}
(\eta + 2) \mu_0 &= \ln (\eta + 1) + \mathcal{O}(\ln \ln (\eta +1)), \quad \eta \rightarrow \infty,
\end{align}
uniformly in $\sigma \in [\epsilon,M]$, where $\epsilon > 0$ and $M > \epsilon$ are fixed.
To show~\eqref{eqn:equation_speed_crit_int2}, we note from $\mu_0 ( 1 + \mu_0)^{\eta + 1} = \sigma$ that
\begin{equation}\label{eqn:equation_speed_crit_int3b}
(\eta + 1) \mu_0 \ge (\eta + 1) \ln(1 + \mu_0) = \ln \sigma - \ln \mu_0.
\end{equation}
Next $\sigma = \mu_0(1 + \mu_0)^{\eta + 1} \ge \mu_0^{\eta + 2}$, and so $\ln \mu_0 \le \frac{1}{\eta + 2} \ln \sigma$. Therefore
\begin{equation}
(\eta + 1) \mu_0 \ge \ln \sigma - \frac{1}{\eta + 2} \ln \sigma = \frac{\eta + 1}{\eta + 2} \ln \sigma,
\end{equation}
and~\eqref{eqn:equation_speed_crit_int2} follows.
As to~\eqref{eqn:equation_speed_crit_int3}, we first observe from~\eqref{eqn:dmu_0_dbeta} that $\mu_0$ decreases in $\eta$ when $\sigma > 0$ is fixed. Hence $L = \lim_{\eta \rightarrow \infty} \mu_0$ exists, and it follows from $\mu_0 (1 + \mu_0)^{\eta +1} = \sigma$ that $L = 0$. Thus, $\mu_0$ decreases to~0 as $\eta \rightarrow \infty$. Then, from~\eqref{eqn:equation_speed_crit_int3b} we get that $(\eta + 1) \mu_0$ increases to $\infty$ as $\eta \rightarrow \infty$. All this holds uniformly in $\sigma \in [\epsilon,M]$: since $\mu_0$ increases in $\sigma$, the right-hand side of~\eqref{eqn:equation_speed_crit_int3b} is bounded below by $\ln \epsilon - \ln \mu_0 (\sigma = M)$. Now take $\eta_0 >0$ such that $(\eta + 1) \mu_0 \ge \sigma$ when $\eta \ge \eta_0$ and $\epsilon \le \sigma \le M$. Then from $\mu_0(1 + \mu_0)^{\eta+1} = \sigma$ we have
\begin{equation}
(\eta + 1) \ln(1 + \mu_0)  = \ln \sigma - \ln \mu_0 \le \ln(\eta + 1) \mu_0 - \ln \mu_0 \le \ln(\eta + 1)
\end{equation}
when $\eta \ge \eta_0$ and $\epsilon \le \sigma \le M$. Hence, when $\eta \ge \eta_0$,
\begin{equation}\label{eqn:equation_speed_crit_int4}
\mu_0 \le {\rm exp}\left[ \frac{\ln(\eta + 1)}{\eta + 1}\right] - 1 = \frac{\ln(\eta + 1)}{\eta + 1} + \mathcal{O}\left( \left( \frac{\ln (\eta + 1)}{\eta + 1} \right)^2 \right),
\end{equation}
where the $\mathcal{O}$ holds uniformly in $\sigma \in [\epsilon, M]$. Then,  by~\eqref{eqn:equation_speed_crit_int3b},
\begin{align}
\nonumber (\eta + 1) \mu_0 &\ge \ln \sigma - \ln \left( {\rm exp} \left[ \frac{\ln (\eta +1)}{\eta + 1} \right] - 1 \right) = \ln \sigma - \ln ( \frac{\ln(\eta + 1)}{\eta+1}\left(1 + \mathcal{O}\left(\frac{\ln(\eta+1)}{\eta+1}\right)\right)\\
\label{eqn:equation_speed_crit_int5} &= \ln (\eta + 1) - \ln \ln (\eta + 1) + \ln \sigma + \mathcal{O}\left( \frac{\ln (\eta + 1)}{\eta + 1} \right),
\end{align}
with $\mathcal{O}$ holding uniformly in $\sigma \in [\epsilon,M]$ and $\eta \ge \eta_0$. From~\eqref{eqn:equation_speed_crit_int4} and~\eqref{eqn:equation_speed_crit_int5} we get~\eqref{eqn:equation_speed_crit_int2} uniformly in $\sigma \in [\epsilon,M]$.


\begin{thebibliography}{10}

\bibitem{Abramson70}
N.~Abramson.
\newblock The {ALOHA} system - another alternative for computer communications.
\newblock In {\em Proc. of AFIPS}, pages 281--285, 1970.

\bibitem{Asmussen03}
S.~Asmussen.
\newblock {\em Applied Probability and Queues}.
\newblock Springer-Verlag, New York, second edition, 2003.

\bibitem{Ba2004}
Y.~Baryshnikov, E.G. Coffman, Jr., and P.~Jelenkovi{\'c}.
\newblock Space filling and depletion.
\newblock {\em Journal of Applied Probability}, 41(3):691--702, 2004.

\bibitem{BoKe80}
R.R. Boorstyn and A.~Kershenbaum.
\newblock Throughput analysis of multihop packet radio.
\newblock In {\em Proc. of ICC}, pages 1361--1366, 1980.

\bibitem{BoMcPr08}
C.~Bordenave, D.~McDonald, and A.~Prouti\`{e}re.
\newblock Performance of random medium access control, an asymptotic approach.
\newblock In {\em Proc. of {ACM Sigmetrics}}, pages 1--12, 2008.

\bibitem{DeBruijn81}
N.G. de~Bruijn.
\newblock {\em Asymptotic methods in analysis}.
\newblock Dover Publications Inc., New York, third edition, 1981.

\bibitem{DeBoVeHi08}
D.~Denteneer, S.C. Borst, P.M. van~de Ven, and G.~Hiertz.
\newblock {IEEE} 802.11s and the philosophers' problem.
\newblock {\em Statistica Neerlandica}, 62(3):283--298, 2008.

\bibitem{DuDoTh07}
M.~Durvy, O.~Dousse, and P.~Thiran.
\newblock Modeling the 802.11 protocol under different capture and sensing
  capabilities.
\newblock In {\em Proc. of INFOCOM}, pages 2356--2360, 2007.

\bibitem{DuDoTh09}
M.~Durvy, O.~Dousse, and P.~Thiran.
\newblock Self-organization properties of {CSMA/CA} systems and their
  consequences on fairness.
\newblock {\em {IEEE} Transactions on Information Theory}, 55(3), 2009.

\bibitem{KlTo75}
L.~Kleinrock and F.A. Tobagi.
\newblock Packet switching in radio channels: part {I} - carrier sense
  multiple-access modes and their throughput-delay characteristics.
\newblock {\em {IEEE} Transactions on Communications}, 23(12):1400--1416, 1975.

\bibitem{LiHo07}
T.Y. Lin and J.C. Hou.
\newblock Interplay of spatial reuse and {SINR}-determined data rates in
  {CSMA/CA}-based, multi-hop, multi-rate wireless networks.
\newblock In {\em Proc. of {INFOCOM}}, pages 803--811, 2007.

\bibitem{MaViRoZh09}
H.~Ma, R.~Vijaykumar, S.~Roy, and J.~Zhu.
\newblock Optimizing 802.11 wireless mesh networks based on physical carrier
  sensing.
\newblock {\em {IEEE/ACM} Transactions on Networking}, 17(5):1550--1563, 2009.

\bibitem{PiYe86}
E.~Pinsky and Y.~Yemini.
\newblock The asymptotic analysis of some packet radio networks.
\newblock {\em {IEEE} Journal on Selected Areas in Communications},
  4(6):938--945, 1986.

\bibitem{RaShSh09}
S.~Rajagopalan, J.~Shin, and D.~Shah.
\newblock Network adiabetic theorem: An efficient randomized protocol for
  contention resolution.
\newblock In {\em Proc. of {ACM Sigmetrics/Performance}}, pages 133--144, 2009.

\bibitem{ToKl75}
F.P. Tobagi and L.~Kleinrock.
\newblock Packet switching in radio channels: part {II} - the hidden terminal
  problem in carrier sense multiple-access and the busy-tone solution.
\newblock {\em {IEEE} Transactions on Communications}, 23(12):1417--1433, 1975.

\bibitem{VeLeDeJa09}
P.M. van~de Ven, J.S.H. van Leeuwaarden, D.~Denteneer, and A.J.E.M.
  Janssen.
\newblock Spatial fairness in wireless multi-access networks.
\newblock In {\em Proc. of {ValueTools}}, pages 51--1/7, 2009.

\bibitem{WaKa05}
X.~Wang and K.~Kar.
\newblock Throughput modelling and fairness issues in {CSMA/CA} based ad-hoc
  networks.
\newblock In {\em Proc. of INFOCOM}, pages 23--34, 2005.

\bibitem{YaVa05}
X.~Yang and N.H. Vaidya.
\newblock On the physical carrier sense in wireless ad hoc networks.
\newblock In {\em Proc. of {INFOCOM}}, pages 2525--2535, 2005.

\bibitem{ZaMo06}
M.~Zafer and E.~Modiano.
\newblock Blocking probability and channel assignment in wireless networks.
\newblock {\em {IEEE} Transactions on Wireless Communications}, 5(4):869--879,
  2006.

\bibitem{ZhFa06}
H.~Zhai and Y.~Fang.
\newblock Physical carrier sensing and spatial reuse in multirate and multihop
  wireless ad hoc networks.
\newblock In {\em Proc. of {INFOCOM}}, pages 1--12, 2006.

\end{thebibliography}
\end{document}